# Assessment of turbulence-chemistry interaction models in MILD combustion regime

ASHOKE DE*, AKSHAY DONGRE

Department of Aerospace Engineering, Indian Institute of Technology, Kanpur, India-208016

**Abstract:** The present paper reports on the assessment of different turbulence-chemistry interaction closures for modeling turbulent combustion in the Moderate and Intense Low oxygen Dilution (MILD) combustion regime. 2D RANS simulations have been carried out to model flames issuing from two burners i.e. Delft-Jet-in-Hot-Coflow (DJHC) burner and Adelaide JHC burner which imitate the MILD combustion. In order to model these flames, two different approaches of turbulence-chemistry interaction models, i.e. Eddy Dissipation Concept (EDC) and PDF based models, are considered; while in the PDF based modeling, two different variants are invoked to understand the applicability of the PDF based models in the MILD regime: one is based on presumed shape PDF (i.e. steady flamelet (SF) model) approach and the other one is transported PDF approach. For transported PDF method, two different solution approaches namely, Lagrangian solution method (LPDF) and Multi-Environment Eulerian PDF (MEPDF) model are considered. A comprehensive study has been carried out by comparing the results obtained from these different models. For the DJHC burner, the computations are carried out for a jet speed corresponding to Reynolds numbers of Re=4100, whereas the Adelaide JHC burner computations are performed for a jet speed corresponding to Reynolds number of Re=10000. The effects of molecular diffusion on the flame characteristics are also studied by using different micro-mixing models. In the case of DJHC burner, it has been observed that the mean axial velocity and the turbulent kinetic energy profiles are in good agreement with the measurements. However, the temperature profiles are over-predicted in the downstream region by both EDC and the PDF based models. In the context of Adelaide JHC burner, the profiles of the temperature and the mass fraction of major species ($CH_4$, $H_2$, $O_2$, $H_2O$, $CO_2$) obtained using LPDF approach are in better agreement with the measurements compared to those obtained using EDC model; although, both the solution approaches fail to capture CO and OH radical profiles.

*Corresponding Author: Tel.: +91-512-2597863 Fax: +91-512-2597561

E-mail address: ashoke@iitk.ac.in

## 1. INTRODUCTION

Owing to the stringent laws placed on pollutant emission in recent years, a major challenge faced by any combustion system designer is the system optimization to reduce the pollutant emission. In this regard, the Moderate and Intense Low oxygen Dilution (MILD) combustion appears to be a promising technique as its key feature lies in the improved thermal efficiency provided the heat to preheat the mixture is extracted from the flue gases. In the MILD combustion, the inlet temperature of the reactants remains higher than the mixture auto-ignition temperature, which is usually obtained by re-circulating the hot product gases into the incoming fresh air. The hot gas re-circulation has two advantages: firstly, it raises the reactant temperature and thereby improving the thermal efficiency of the system through heat recovery and secondly, it produces diluted mixture (low $O_2$ concentration), which reduces the $NO_x$ emissions owing to a lower flame temperature. Few other important features of the MILD combustion include flat radiation flux, low turbulence fluctuations, flat temperature field, barely visible and audible flame. These remarkable features are the reasons why MILD combustion has received attention in both experimental [1-4] and numerical [5-18] studies.

One of the common configurations used for achieving MILD combustion is the Jet-in-Hot-Coflow (JHC) flame, where the reaction takes place in the presence of hot oxidizer stream containing diluted combustion products. Both the burners, i.e. DJHC and Adelaide JHC, studied herein fall under the same category. The DJHC burner was designed in the Delft University of Technology (TU-DELFT) with the objective of studying the basics of MILD combustion. Oldenhof et al. [1] studied the JHC flame stabilization in the MILD regime by recording the flame luminescence with the help of an intensified high-speed camera. They studied the flame lift-off height by studying the formation of ignition kernels which convect downstream to form a stable flame. They found that an increase in temperature of the coflow and addition of higher alkanes lowers the flame lift-off height. Also, an increase in jet Reynolds number led to a significant decrease in flame lift-off height. But increasing the Reynolds number beyond a certain limit affected the growth of flame pockets, thus increasing the lift-off height. In a further study, Oldenhof et al. [2] reported the role of entrainment in JHC flame stabilization in the MILD regime. The principle focus of their study was to understand the turbulence-chemistry interaction in the MILD regime. They found that the mean velocity and turbulent stresses were not affected by the chemical reactions. They also reported that the reduction in the flame lift-off height at higher Reynolds number is due to an increase in jet entrainment at higher Reynolds number in conjunction with the positive temperature gradient in radial direction near the jet exit area. In Oldenhof et al. [3], the authors tried to study the characteristics of ignition of impulsively started jets of natural gas in a hot coflow with the primary goal of understanding the processes that occur in a furnace with regenerative burners. The series of experiments performed by Oldenhof et al. [1-3] provided a detailed database but their database had only velocity, temperature and turbulent quantities. The Adelaide JHC burner [4], designed in the Adelaide University, provided a detailed database for species in the MILD regime. Dally et al. [4] studied the flame characteristics of turbulent non-premixed jet flames in a diluted hot coflow using the Adelaide JHC burner. Both temporally and spatially resolved measurements of reactive scalars were performed on methane-hydrogen flames with different oxygen levels in the hot oxidizer stream at a fixed jet Reynolds number. Substantial variations in the flame structure and characteristics were observed by decreasing the oxygen level. This experiment provided species data for major as well as minor species but lacked data on turbulent quantities and velocity profiles. Thus, the experiments performed on both Delft and Adelaide JHC burners jointly provided a detailed database to study the JHC flames and has been used in the present study to understand the applicability of various combustion models in the MILD regime.

Flames issuing from both the burners were subjected to different numerical studies. In context of the DJHC burner, De et al. [5] studied the effects of different turbulence models, like Standard k-ε (SKE), Realizable k-ε (RKE), and Renormalization Group k-ε (RNG) model, in conjunction with Eddy Dissipation Concept (EDC) turbulence-chemistry interaction model. They observed that for the higher Reynolds number case (Re=8800), RKE turbulence model produced better results than the SKE model whereas for the lower Reynolds number case (Re=4100), RKE produced agreeable profiles for mean velocity but under-predicted the turbulent kinetic energy profiles significantly. They also concluded that in order to capture the flame characteristics in MILD regime, the chemical time scale must be controlled properly. Recently, De et al. [6]

reported the DJHC flames using the transported PDF models i.e. Lagrangian PDF (LPDF) and MEPDF model and found that both models produced similar results which were found to be in good agreement with the experimental measurements excluding some discrepancies in the downstream region. Major sources of errors in modeling these flames are the inaccuracies in modeling turbulence-chemistry interaction which were reported in this paper. Kulkarni et al. [7] used Large Eddy Simulation (LES) method in conjunction with stochastic fields method and tabulated chemistry to study the DJHC flames in order to understand the role of entrainment on flame stabilization and effects of Reynolds number on the lift-off height. They observed that LES predicted the lift-off height reasonably well for different Reynolds number but temperature profiles showed discrepancies between the predictions and measurements. In a recent study, Bhaya et al. [8] used the LES methodology along with transported PDF combustion models to simulate the MILD combustion regime. Their study included both the Eulerian and Lagrangian approaches to study the DJHC flames. They found that both the PDF models predicted the mean velocity and turbulent kinetic energy profiles accurately whereas an early ignition resulted in higher mean temperature.

In the context of the Adelaide burner, Christo et al. [9] studied methane-hydrogen flames using EDC combustion model in conjunction with different k-ε models (Standard k-ε model, Realizable k-ε model, and Renormalization group k-ε model) and detailed chemistry in order to evaluate the performance of turbulence and combustion model along with different chemical mechanisms. The major finding of their study was that the SKE turbulence model with a modified dissipation equation constant ($C_{\varepsilon 1}$=1.6) produced the best agreement with the experiments. They also observed that differential diffusion effects have a strong impact on the flame characteristics. In a further study, Christo et al. [10] modeled these flames using a hybrid

RANS/PDF approach in combination with different chemical mechanisms (Smooke [19], ARM, and GRI 3.0 [20]) and Euclidean Minimum Spanning Tree (EMST) [21] micro-mixing model. The interesting and major finding of their study was that the use of sophisticated high-temperature/high-oxygen optimized chemical mechanisms won't always improve the accuracy but may adversely affect the results in a way that increases the sensitivity of the solution to slight variations in flow conditions. Later on, Kim et al. [11] modeled these flames using Conditional Moment Closure (CMC) approach. The primary focus of their study was to study the JHC flame structure and characteristics along with NO formation in the MILD regime. There are several other RANS based modeling studies carried out on these flames with most notable of them being the studies by Frassoldati et al. [12], Mardani et al. [13, 14] and Aminian et al. [15]. All of these studies included the use of EDC combustion model in combination with DRM 22 [22], GRI 2.11 [23] and KEE-58 [24] chemical mechanisms with the primary goal of understanding the flame structure and molecular diffusion in the MILD regime. On the other hand, Ihme et al. [16, 17] modeled these flames using the LES methodology by considering the burner as a three-stream problem. In order to capture the third stream, they introduced an additional conserved scalar using Flamelet/Progress Variable (FPV) formulation. Recently, Dongre et al. [18] modeled JHC flames in both DJHC as well as Adelaide burner using the MEPDF-IEM methodology with the primary goal of assessing its accuracy and predictive capabilities in the MILD regime. They considered effects of differential diffusion as well as chemistry on the flame characteristics and reported that using detailed or reduced chemical mechanisms had no substantial effects on the flame characteristics in the MILD regime.

In the MILD regime, reaction rates are relatively slow compared to a simple non-premixed combustion of fuel and air, thereby enhancing the influence of molecular diffusion on flame characteristics. Due to these two combined effects, the MILD combustion modeling offers a great challenge to the various popular combustion modeling approaches. For example, the steady flamelet model is more suitable to fast chemistry while the PDF transport approaches have the molecular mixing as one the major source of errors.

Since none of the available literatures reported the performance analysis of different variants of turbulence-chemistry interaction models together (SF, MEPDF, LPDF, EDC) in MILD regime, the primary goal of the present work is to assess the performance of both PDF based (both presumed shape PDF and transported PDF) approaches along with finite rate kinetics based (EDC) combustion models and bring further insight to understand their accuracy and applicability to model the flames in MILD combustion regime. A comprehensive study has been carried out by comparing the results obtained using both presumed shape PDF based model (SF) and transported PDF based (LPDF, MEPDF) models with those obtained using EDC combustion model, which none of the previous studies reported. The numerical predictions obtained using these models are compared with respective experimental databases for the corresponding Reynolds number cases i.e. Re=4100 for the DJHC burner and Re=10000 for the Adelaide Burner.

## 2. TURBULENCE-CHEMISTRY INTERACTION MODELS

### 2.1 PDF based models

PDF based turbulence-chemistry interaction models are based on stochastic method, which directly considers the probability distribution of the relevant stochastic quantities in a turbulent reacting flow. These models are categorized into two classes: one is presumed shape PDF modelling and the other is transported PDF modelling. Both the approaches are briefly discussed below.

**2.1.1 Presumed shape PDF approach**

In presumed PDF approach, which essentially assumes the shape of the probability function P, is simpler to use. The turbulent flow field is defined using the function P and the properties obtained from laminar flamelet model, which considers the turbulent flame as an ensemble of laminar and one dimensional local structures. The flame surface is defined as an iso-surface of the mixture fraction within the turbulent flow field. Counter flow configuration of laminar diffusion flame is used to represent the thin reactive-diffusive layers in the turbulent flow field. The flame equations can be transformed from physical to mixture fraction space to represent the reactive-diffusive layer. The theoretical description and formulation of the steady flamelet model can be referred from Peters [25, 26].

**2.1.2 Transported PDF models**

The starting point of the Probability Density Function (PDF) transport modeling is the joint composition PDF transport equation which is given below:

$$\frac{\partial \rho f_\varphi}{\partial t} + \frac{\partial}{\partial x_i}\left[\rho u_i f_\varphi\right] + \frac{\partial}{\partial \psi_k}\left[\rho S_k f_\varphi\right] = -\frac{\partial}{\partial x_i}\left[\rho \langle u_i'' | \psi \rangle f_\varphi\right] + \frac{\partial}{\partial \psi_k}\left[\rho \left\langle \frac{1}{\rho}\frac{\partial J_{i,k}}{\partial x_i} | \psi \right\rangle f_\varphi\right] \qquad (1)$$

Where $f_\phi$ is the single-point, joint probability density function (PDF) of species composition and enthalpy. The first term on the left hand side of eqn. (1) represents the unsteady rate of change of PDF while the second term is the transport in physical space by mean velocity. Finally, the third term represents the transport in composition space by chemical reactions. On the right hand side, the first and second term represents transport due to velocity fluctuation, which is closed using a turbulence model, and molecular mixing, which is closed using a micro-mixing model, respectively. The PDF formulation allows the chemical source term in closed form (Eqn. 1) along with calculation of scalar fluctuations which gives PDF models an advantage over other combustion models while simulating turbulent non-premixed flames as reported by Jaishree et al. [27] and Rakesh et al. [28, 29]. The approach used to discretize and solve eqn. (1) distinguishes the Lagrangian method from the Eulerian one. A complete description on the transported PDF methods can be found in Pope [30] and Fox [31]. A complete review on transported PDF methods can be found in Haworth [32].

**2.1.2.1 Lagrangian PDF method (LPDF)**

In the Lagrangian method [30], the turbulent reacting flow is modeled by a system of 'notional' particles whose one-point, one-time Eulerian joint PDF evolves as per eqn. (1). The particles move stochastically in the physical space and carry mass. The total mass of the particles is equal to the total mass of the control volume. During iteration, the states of the particles change due to convection, mixing, and reaction and, in our implementation, these processes are tracked in fractional steps.

In the composition PDF formulation, the notional particles do not carry velocity and the mass density function is defined as:

$$F^*_{\phi x}(\psi, y; t) \equiv \sum_{i=1}^{N_p} m^{(i)} \delta\left(\psi - \phi^{(i)}(t)\right) \delta\left(y - x^{(i)}(t)\right) \tag{2}$$

where $\delta\left(y - x^{(i)}(t)\right)$ is a three dimensional delta function at the particle location and similarly for the scalar is $\delta\left(\psi - \phi^{(i)}(t)\right)$, where $F^*$ is expressed as:

$$\left\langle F_{\phi x}^{*}(V,\psi,y;t)\right\rangle = \left\langle \rho(x,t)\right\rangle \tilde{f}_{\phi}(\psi;x,t) = \rho(\psi) f_{\phi}(\psi;x,t) \tag{3}$$

where $\tilde{f}_{\phi}(\psi,x;t)$ represents the Favre PDF $\tilde{f}_{\phi}$, $f_{\phi}(\psi,x;t)$ represents the conventional PDF $f_{\phi}$ and $\left\langle \rho \right\rangle$ is the mean mixture mass density.

For an infinitesimal time increment d$t$, the position, and composition of each notional particle evolves as;

$$dx_{i}^{*} = \tilde{u}_{i}^{*} dt + dx_{i,turb}^{*} \qquad (i=1,2,3) \tag{4}$$

$$d\phi_{k}^{*} = S_{k}(\phi^{*})dt + \theta_{k,mix}^{*} dt \qquad (k=1,2,......,N_{\phi}) \tag{5}$$

In this system, the i$^{th}$ particle is assigned a mass $m^{(i)}$ and each particle is distinguished by position co-ordinates $x^{(i)}(t)$ and $N_{\varphi}$ scalar variables $\varphi^{(i)}(t)$. In the above system of equations, $dx_{turb}^{*}$ refers to the increment in particle position due to turbulent velocity fluctuations (turbulent diffusion) about the local mean velocity $\tilde{u}_{i}^{*}$ and usually is modeled using a random-walk model. The random walk model for stochastic particle motion depends on turbulent Schmidt number whose value has been fixed at 0.7 in the present LPDF simulations. The increment in the composition is due to the chemical source term $S_{k}$ and molecular diffusion $\theta_{k,mix}^{*}$, i.e. micromixing. In general, the largest modelling error comes from the molecular or micro-mixing term. Thus, in order to capture the micro-mixing of the species, we have used three different mixing models; Interaction-by-Exchange-with-the-Mean (IEM) [33], Coalescence Dispersion (CD) [34], and Euclidean Minimum Spanning Tree (EMST) [21] model.

## 2.1.2.2 Multi-Environment Eulerian PDF (MEPDF) method

Fox was the first to put forth this method and the details on the derivations of the MEPDF method along with underlying assumptions can be found in his book [31]. In this approach, the joint composition PDF transport equation (eqn. 1) is approximated to have a shape having a series of Delta functions with fixed or variable probability. Thus, the joint composition PDF equation can be represented as a series of $N_e$ Delta functions as follows;

$$f_\phi(\psi;x,t) = \sum_{n=1}^{N_e} w_n(x,t) \prod_{\alpha=1}^{N_s} \delta[\psi_\alpha - <\phi_\alpha>_n (x,t)] \qquad (6)$$

where $N_s$ is the number of species, $N_e$ is the number of environments, $w_n$ is the weight (or probability) of each environment, and $<\phi_\alpha>_n$ is the mean composition vector in the $n^{th}$ environment. Using the definition of eqn. (6), the joint composition PDF eqn. (1) can be transformed into the following set of equations;

$$\frac{\partial \rho p_n}{\partial t} + \frac{\partial}{\partial x_i}(\rho u_i p_n) = \Gamma \nabla^2 p_n \qquad (7)$$

$$\frac{\partial \rho \vec{s_n}}{\partial t} + \frac{\partial}{\partial x_i}(\rho u_i \vec{s_n}) = \Gamma \nabla^2 \vec{s_n} + C_\phi \frac{\varepsilon}{k}(\langle \phi_i \rangle - \langle \phi_i \rangle_n) + p_n S(\langle \vec{\phi_n} \rangle) + \vec{b_n} \qquad (8)$$

where, $\vec{s_n} = p_n \vec{Y_n}$ or $p_n H_n$

The transport equation for the probability of occurrence of $n^{th}$ environment is represented by eqn. (7) whereas the transport equation for the probability of weighted species mass fraction of probability weighted enthalpy in each environment is represented by eqn. (8). In eqn. (8), the second term on the right hand side represents micro-mixing term which is closed by the IEM micro-mixing model whereas the third and fourth term represents reaction source term and correction term, which accounts for modeling assumptions used in MEPDF model. These correction terms are approximated as;

$$\sum_{n=1}^{N_e} \langle \phi \rangle_n^{m_j-1} b_n = \sum_{n=1}^{N_e} (m_j-1) \langle \phi \rangle_n^{m_j-2} p_n c_n \qquad (9)$$

These transport equations are discretized using the finite volume method and solved for each environment. A stiff ODE solver is used to evaluate the reaction source term for each environment and In-Situ-Adaptive-Tabulation (ISAT) [35] is used to tabulate those reaction source terms.

The DQMOM method is applied to derive (eqn. 9) the correction terms. The derivation begins by substituting eqn. (6) into eqn. (1) and following the derivation explained in Fox [31], we obtain a set of linear equations, for $N_s$ scalars, given by eqn. (7) and eqn. (8). The $N_e(N_s+1)$ unknowns are calculated by solving eqn. (8) with lower order moments. The zero[th] and first order moments must be satisfied and yield $(N_s+1)$ linear equations whereas the remaining $(N_e-1)(N_s+1)$ linear equations must be calculated from higher order moments. The term '$m_j$' in eqn. (9) represents the order of moment whose value varies as $m_j=1, 2,…N_e$, where $N_e$ represents the number of environments whereas $N_s$ represents the number of scalars considered.

This method retains the advantages of PDF transport model as the reaction source term appears in the closed form in the transport equations which are then solved in the Eulerian frame but the important fact which makes the difference in this approach is that this method is free from the statistical errors. Jaishree et al. [27] carried out simulation of Sandia flames D, E, and F using the MEPDF model with 2-environment formulation and reproducing only first and second order moments. They mentioned that a small number of environments may not be sufficient enough to obtain good predictions, however, based on studies carried out by Rakesh et al. [28,29], Tang et al.[36], Akroyd et al.[37,38], and Bhaya et al. [8] we can also highlight the fact that a small number of environments is sufficient for reasonable accuracy to predict temperature, velocity, and turbulent quantities which makes this method computationally efficient.

In the MEPDF model, there are certain numerical aspects related to bounded-ness and singularity of the co-variance matrix of different scalars while calculating the generic solutions to find correction terms as reported by Fox and Wang [39]. These issues are handled by imposing constraints like handling each scalar independently and ignoring cross moments. The addition of these constraints creates another issue as the correction terms calculated, then, do not represent the exact moments of the PDF transport equation (eqn. 1) which is another possible source of errors other than the IEM closure in the MEPDF predictions. In addition, Jaishree et al. [27] pointed out that in case of MEPDF model, the source terms are ill-conditioned and realizability is violated for multi-component systems. They also pointed out the ambiguities in MEPDF boundary condition specification and their physical interpretation hampers their accuracy.

## 2.2 Eddy Dissipation Concept (EDC) combustion model

The Eddy Dissipation Concept combustion model was developed by Magnussen et al. [40, 41]. The major advantage of the EDC model is that it is capable of integrating the finite rate kinetics effects at the expense of computational cost which makes EDC an attractive choice to model JHC flames but this advantage compromises the model's ability to describe turbulent temperature fluctuations.

In the EDC model, the computational cell is theoretically partitioned into two sub-zones: 'fine structure', where all the homogeneous chemical reactions are assumed to occur, and 'the surrounding fluid' where only turbulent mixing takes place without any chemical reaction. These chemical reactions are treated locally as adiabatic, isobaric, Perfectly Stirred Reactors (PSR) which transfer the mass and energy only to the surrounding fluid, thus, transporting surrounding reactant and product gases to and from the fine structure. With this assumption, the reaction rates of all the species are calculated from a mass balance of the fine structure reactor. The mean species reaction rate for the transport equation is given as follows;

$$\tilde{\omega}_i = \frac{\overline{\rho}\xi^2\lambda}{\tau}\left(Y_i^* - Y_i^0\right) \tag{10}$$

where $Y_i^*$ and $Y_i^0$ represent the mass fraction of species, $i$, in the reacting and non-reacting part, respectively. $\tau$, which represents the mean residence time, is the inverse of the specific mass exchange rate between the fine structures and the surrounding. $\lambda$ represents the fraction of the fine structure where the reaction takes place.

The mean mass fraction $\tilde{Y}_i$ is evaluated using the linear combination of properties in the fine structures and the surrounding fluid as;

$$\tilde{Y}_i = \xi^3\lambda Y_i^* + \left(1 - \xi^3\lambda\right)Y_i^0 \tag{11}$$

The choice of $\lambda=1$ serves best when using EDC with detailed chemistry as mentioned in De et al. [5] after performing a sensitivity analysis. Substituting this value in the above equation, we can evaluate the value of $Y_i^0$ as follows:

$$Y_i^0 = \frac{\tilde{Y}_i - \xi^3 Y_i^*}{\left(1 - \xi^3\right)} \tag{12}$$

Rearranging the eqns. (10) and (12), we obtain the expression for mean chemical source term as follows:

$$\tilde{\dot{\omega}}_i = \frac{\overline{\rho}\xi^2}{\tau\left(1-\xi^3\right)}\left(Y_i^* - \tilde{Y}_i\right) \tag{13}$$

In EDC, the combustion in the fine structure is presumed to occur in a constant pressure reactor instead of a PSR using the initial conditions taken from that of current species and temperature in the cell. In this approach, which is called as Plug Flow Reactor (PFR), reactions proceed over the time scale, $\tau$, given by Arrhenius rates and are integrated numerically in time. In the present study, we have used the CFD code Fluent [42] which uses the PFR approach.

In the EDC model, the size of fine structure $\xi$ and the mean residence time $\tau$ is evaluated as;

$$\xi = C_\xi \left(\frac{\nu\varepsilon}{k^2}\right)^{1/4} \tag{14}$$

and

$$\tau = C_\tau \left(\frac{\nu}{\varepsilon}\right)^{1/2} \tag{15}$$

Here, the model constants have the default values of: $C_\xi = 2.1377, C_\tau = 0.4082$ [5]. Since, EDC formulation does not include calculation of turbulent fluctuations, an extra scalar equation has been solved using a user defined function to calculate temperature fluctuations as detailed in De et al. [5].

## 3. DELFT-JET-IN-HOT-COFLOW (DJHC) BURNER

In this section, the numerical predictions obtained using SF, EDC, LPDF, and MEPDF models, for the DJHC burner are reported.

### 3.1 Burner description and numerical setup

The Delft-Jet-in-Hot-Coflow burner was developed in TU Delft by Oldenhof et al. [1-3] to study the MILD combustion. The burner consists of a central fuel jet pipe with an internal diameter of 4.5 mm. Surrounding this fuel jet pipe is an outer tube of internal diameter 82.8 mm, which houses a secondary burner that helps to generate a hot coflow. This secondary burner is a ring burner consisting of a ring of premixed flame while fresh air is injected on both sides of the ring burner, thus it operates in partially premixed combustion mode.

Considering the symmetry of the burner, a 2D axi-symmetric grid has been used to perform the simulations. In the experiments, the first LDA measurement was taken at z=3 mm from the fuel exit. Considering this, the computational domain starts 3 mm downstream of the jet exit and extends for 225 mm in the axial direction while it extends 80 mm in radial direction. The simulations are performed using ANSYS Fluent 13.0 [42]. The simulations have been carried out with Dutch Natural gas as the fuel. The composition of the Dutch natural gas in the present formulation is 81% $CH_4$, 4% $C_2H_6$, and 15% $N_2$ whereas the hot coflow composition contains the combustion products of a ring burner as mentioned in the experimental papers [1-3], which contains $O_2$ with a constant percentage of 7.6 % (by mass). The Reynolds number, based on the fuel jet, is 4100, therefore the flow is turbulent and modeled using RANS based Realizable k-ε (RKE) turbulence model. The chemistry is modeled using a reduced mechanism, DRM19 [22], which include 19 species and 84 reactions. In context of the MEPDF model, IEM micro-mixing model has been used to close the mixing term in eqn. (9) whereas for the LPDF model, three different mixing models namely CD, EMST, and IEM, are used. The pressure and velocity coupling has been done using SIMPLE algorithm and a second order upwind scheme has been used to compute the convective fluxes of all the equations. The mean velocity and mean temperature profiles used at the inlet boundaries of both fuel jet and coflow are referred from the experimental database [1]. The experimental radial profiles at x=3 mm have been used to define boundary conditions of the physical quantities at fuel and coflow inlets. Since, the experimental database for the DJHC burner did not include the species data, the species concentrations are calculated using the equilibrium assumptions as provided in [5]. The turbulent kinetic energy and dissipation are calculated from the measured axial and radial normal components of the Reynolds stresses by assuming the azimuthal component to be equal to the radial component. For the transported PDF model, the micro-mixing model constant is kept 2.0. In particular, the computations using MEPDF model are performed with two environments whereas for the LPDF model, the computations are performed by taking 30 particles per cell to reduce the statistical errors. The value of turbulent Schmidt number, used in the random walk model, has been taken as 0.7. It is also noteworthy that the temperature fluctuations with RMS value 100K that are observed experimentally have not been taken into account in modelling. Detailed description on the burner geometry and numerical setup can be referred from De et al. [5]. In order to study how the predictions perform against the measurements, the measurements from both the sides of symmetry are also plotted in the same frame. For any radial plots, the radial range of these measurements is given as: *triangle* $0 \leq r \leq 35$, *squares* $-35 \leq r \leq 0$.

## 3.2 Results and Discussion

In this section, we discuss the results obtained from simulations performed for the DJHC burner, exhibiting the coflow properties explained in Oldenhof et al. [1], at Reynolds number Re=4100. The obtained results are compared with the experimental database provided by Oldenhof et al. [1]. The oxygen concentration has been kept constant at 7.6% (by mass) for DJHC-I flame. A grid with 180x125 (axial x radial) cells has been used to perform simulations after performing a grid independence study shown in Figure 1. First, the results for mean axial velocity, turbulent kinetic energy and mean temperature are discussed followed by the predictions of Reynolds stresses for the lower Reynolds number case (Re=4100).

Figure 2 depicts the radial profiles of mean axial velocity, turbulent kinetic energy and mean temperature obtained for DJHC-I flame (Re=4100) with different combustion models. As observed from the figure, the predictions show a proper trend of mean axial velocity and are in good agreement with the experimental measurements for all the approaches. Similarly, the turbulent kinetic energy profiles are properly predicted by SF, EDC, MEPDF, and LPDF models with slight under-estimation near the jet exit and a slight over-prediction in the downstream region. These discrepancies are due to the inaccuracies in the turbulence model which is unable to predict the production and dissipation of the turbulent kinetic energy properly. Major discrepancies between the predictions and measurements for different turbulence-chemistry interaction models can be observed in the mean temperature profiles. Initially, near the jet exit area (till x=30 mm), all the four combustion models predict the temperature profiles well but in the downstream region the discrepancies start to appear. The peak temperature is observed at axial location x=150 mm from the jet exit at r~17 mm. The peak temperature predicted by steady flamelet model is 1540 K while for the EDC model, the predicted peak temperature is 1825 K, which is around 25% higher than the experimental measurements and is in good agreement with the results obtained by De et al. [5]. Whereas in the case of transported PDF models, the peak temperature predicted by the Lagrangian PDF models is 1660 K, 1655K and 1730 K with EMST, CD and IEM closures, respectively, while the MEPDF model shows 1750 K, which is around 13% higher than the measurements [1]. As noted earlier, the limitations of SF formulation exhibit the discrepancies in mean temperature and turbulent kinetic energy profiles. Similarly, EDC model performs well in fuel lean region, but fails to capture the profiles in fuel rich region. While comparing the transported PDF predictions, we notice that the peak temperature is observed at axial location x=150 mm and radial location r=10 mm with EMST, whereas in case of CD and IEM it shifts radially to 12 mm and 8 mm radial locations, respectively. Also the difference in peak temperature between these mixing models is more than 100K which emphasizes the role of micro-mixing here. On the other hand, in case of the MEPDF model, the early ignition peaks are also observed at the axial location of x=60mm.

It is essential to investigate the Reynolds stresses, obtained using the Boussinesq hypothesis, with different turbulence-chemistry interaction models as they describe the mixing of momentum between fuel jet and hot coflow. Figure 3 depicts the radial profiles of the Reynolds stresses obtained for the DJHC-I (Re=4100) flame. As observed from the figure, all the models under-predict the normal stress component (u'u') till the axial location x=30mm after which they are properly captured in the downstream region. The Reynolds shear stress component (u'v') is accurately captured by all the models and is in good agreement with the experimental measurements. Therefore, the discrepancies observed while predicting the Reynolds stresses can be primarily attributed to the inaccuracies of the turbulence models.

Oldenhof et al. [1] introduced the probability of burning based on detection of ignition kernel as a method to measure the lift-off height. The ignition kernels were detected in the chemi-luminescence signal which was based on the excited state of OH. The threshold for detection followed the settings of the camera. In the present study, we have used mean OH mass fraction to determine the lift-off height with a threshold value of 1e-3 [7-8]. The SF model is unable to capture the delay in chemical reaction and therefore gives a flame which is attached to the burner. Other than the SF, all other models predicted a lift-off height. The flame lift-off height captured by EDC and MEPDF-IEM models is 41 mm and 33 mm, respectively, whereas that captured by LPDF-IEM, LPDF-EMST, and LPDF-CD models is 36 mm, 39 mm, and 40 mm, respectively. Although EDC and LPDF models produce better results compared to the MEPDF model, all of them under-predicted the lift off height compared to the reported experimental lift-off height which is close to 84 mm [1]. Based on the results obtained for the mean quantities, we still cannot distinguish the performance of different models as the predictions are of similar nature. Therefore, in order to understand the predictive capabilities of these models, it is worthwhile to analyze their performances against the RMS quantities; thus, we study the RMS profiles of temperature as depicted in Figure 4. Differences among predictions can be noted here, where the SF and MEPDF models fail to capture the temperature RMS profiles completely. Whereas, on the other hand, having solved an extra scalar equation for temperature RMS results in EDC, the profiles are captured accurately near the jet exit but as we move downstream it slightly over-predicts them. In the present simulations, the measured temperature fluctuations have not been added at the inflow boundary conditions in the LPDF models and its effect is evident from the figure as the temperature RMS profiles are significantly under-predicted at the centerline near the jet exit but as we move downstream, the profiles with LPDF models improve adequately along centerline. Away from the centerline, the LPDF models capture the trend of measurements but are significantly under-predicted. Therefore, it must be noted that, in case of the LPDF models, providing only mean temperature at the inflow boundary conditions is not sufficient and fluctuations must also be included to obtain better predictions.

In order to be more conclusive, it is important to assess these models against mean as well as RMS species profiles; as the DJHC burner does not provide any species data, it is worthwhile to continue this exercise for the Adelaide burner which provides a detailed species data which will help us to gain detailed understanding of the validity of these models in the MILD regime.

## 4. ADELAIDE JHC BURNER

In this section, we report the numerical predictions obtained, using the SF, EDC, LPDF, and MEPDF models for the Adelaide burner.

### 4.1 Burner Description and Numerical Set-up

The construction of the Adelaide burner is very similar to the DJHC burner. The main difference between the two lies in the cooling method used to cool down the coflow. In case of the Adelaide burner, $N_2$ is used for cooling down the coflow while in case of the DJHC burner, the coflow is cooled through convective and radiative heat losses that take place along the burner pipe. The second remarkable difference between the two burners is in the design of the secondary burner. The design of the secondary burner of the DJHC burner allows for particle seeding, used in LDA and PIV measurements, while the Adelaide burner has no such arrangements. Another difference between the burners is that Adelaide burner uses porous ceramic straps to minimize the heat loss to the surroundings whereas DJHC burner uses a perforated distribution plate which promotes heat loss by radiation from the heated plate. The central fuel jet pipe of the Adelaide burner has an internal diameter of 4.25 mm, which is surrounded by an outer annulus of internal diameter 82 mm. Detailed description on the burner schematics can be referred from Dally et al. [4].

In order to model the methane-hydrogen JHC flames in the Adelaide burner, a numerical setup similar to the DJHC burner case, though with a few changes, has been used. Three different flames with different oxygen content in the hot coflow have been simulated, namely HM1 (3% $O_2$), HM2 (6% $O_2$), and HM3 (9% $O_2$). In this case also, 2D axisymmetric grid is used for simulations where the computational domain extends for 300 mm X 80 mm in the axial and radial directions, respectively. A modified Standard k-ε (SKE) turbulence model ($C_{\varepsilon 1}$=1.6) has been used to model turbulence while DRM19 [22] is invoked to model the chemistry. The Reynolds number has been kept constant at Re=10000 for all the three flames simulated here. The radial profiles of the temperature and species mass fraction with experimental values at x= 4 mm have been used to specify boundary conditions at fuel jet and coflow inlets. The outlet boundary condition is set to outflow. For the fuel jet inlet, fully developed turbulent pipe flow profiles are used to specify the velocity. For hot and cold coflow inlets, the constant velocity boundary conditions are set to 3.2 m/s and 3.3 m/s, respectively [4]. The studies carried out by Frassoldati et al. [12] showed that the solution is very sensitive to the turbulent quantities at the inlets. Therefore, the turbulent intensities at the hot and cold coflow have been set to 5% while it is set to 7% at the fuel jet inlet following the recommendations of Frassoldati et al. [12]. Similar to DJHC case, the measurements from both the sides of symmetry are also plotted in the same frame to study how the predictions perform against the measurements. For any radial plots, the radial range of these measurements is given as: *triangle* $0 \leq r \leq 70$, *squares* $-25 \leq r \leq 0$.

## 4.2 Results and Discussion

In this section, we present the results obtained for the simulations performed for Adelaide JHC flames namely, HM1, HM2 and HM3, exhibiting the coflow properties explained in Dally et al.[4] at Reynolds number Re=10000. The grid independence study has been carried out using two grids; a coarse grid with 400x130 (axial x radial) cells and a fine grid with 800x240 cells. Figure 5 shows the results of the grid independence study carried out for the HM1 flame using above mentioned grids using the MEPDF combustion model along with modified SKE turbulence model and DRM19 [22] chemical mechanism. As observed from the figure, the radial as well as centerline mean mixture fraction and RMS mixture fraction profiles obtained using both the grids are in good agreement with the measurements as well as with each other. Therefore, the coarse grid with 400x130 cells has been chosen for the detailed simulations.

Previously, in the case of Delft burner, it is observed that the RMS temperature predictions obtained using all the models are in reasonable agreement with the measurements except SF and MEPDF models, which prompted us to assess the predictive capability of SF and MEPDF models to capture the species profiles at the beginning as this is not possible with DJHC case due to un-availability of species data. Figure 6 depicts the radial profiles of mean $CO_2$, $H_2O$, CO and OH mass fraction obtained for HM1 flame at two axial locations (x=60 mm and x=120 mm) using these two models, i.e. SF and MEPDF. As observed, both the models significantly over-predict the $CO_2$ and $H_2O$ mass fraction profiles whereas in case of CO and OH profiles, SF over-predicts the profiles while MEPDF under-predicts them. The SF model involves a selection of fuel and oxidizer streams; while in the present work, air is selected as oxidizer stream as a result of which coflow has to be presented as a product stream, corresponding to an intermediate value of mixture fraction, at the lean side of stoichiometry. From this configuration, it is clear that the steady flamelets based on air are not representative for the phenomenon close to the burner where mixing between fuel and coflow is intense. In particular, the implicit assumption that fluctuations can be described by beta function PDF for fluctuations over the whole range from 0 (air) to 1 (fuel) is not representative. Christo et al. [9] have already pointed out the deficiencies of the SF model to simulate JHC configuration. Instead of a single mixture fraction formulation, two mixture fractions can be used to overcome these problems as mentioned by Etaati et al. [43] and Ihme et al. [16, 17]. Similarly, Jaishree et al. [27] have pointed out the shortcomings of the MEPDF model to simulate pilot stabilized flames which are already mentioned in section 2.1.2.2. These errors and inaccuracies in the MEPDF formulation are responsible for such bad predictions. Therefore, the results obtained using SF and MEPDF models have been discarded here and only the predictions obtained using the EDC and LPDF models, with DRM19 [22] chemical mechanism and modified SKE turbulence model, will be discussed in the remaining section.

In order to check whether the LPDF results are statistically converged or not, we carried out LPDF-EMST simulations using 20, 30, and 50 particles per cell for HM1 flame which are depicted in Figures 7 and 8. As observed, the mean and RMS profiles of $H_2O$ and $O_2$ mass fractions do not have any significant differences. Therefore, we have considered 30 particles per cell for the rest of simulations.

Figure 9 shows the radial profiles of mean mixture fraction obtained for HM1, HM2, and HM3 flames. As observed, LPDF and EDC models capture the profiles accurately and no significant differences are observed amongst these predictions. Major discrepancies can be observed in the mean temperature profiles as depicted in Figure 10 for these three different flames. As we observe, all the predictions are similar in nature except some improvements obtained using EDC model. None of the models are able to properly capture the centerline temperature which remains significantly under-predicted for all three flames; however, the predictions are improved along the shear layer. Notably, only EDC model is able to predict the peak temperature nicely for all three flames despite the fact that the PDF models incorporate the effects of temperature fluctuations in its formulations. Among the transported PDF models, the LPDF-EMST predictions are better compared to the LPDF-IEM and LPDF-CD predictions. The peak temperature for EDC, in all three flames, is obtained at axial location x=120 mm at an approximate radial location of r=18 mm. In case of HM1 flame, the peak temperature obtained with the EDC model is close to 1400K. Comparing the EDC results obtained with those obtained by Mardani et al. [12], the EDC results show a similar peak temperature. On the other hand, the peak temperature obtained with LPDF models is approximately 1300K which is obtained at the axial location x=60 mm. In case of HM2 flame, the peak temperature, for EDC as well as LPDF models, is obtained at axial location x=120 mm at an approximate radial location of r=20 mm. The peak temperature obtained with EDC is approximately 1600K whereas that obtained using the LPDF-EMST and LPDF-IEM models are close to 1400K. The peak temperature obtained with LPDF-CD is lower compared to other models with a peak temperature of 1160K. In case of HM3 flame, the peak temperature obtained using the EDC model is 1800K at axial location x=120 mm which is approximately 20% higher than the experimental measurements. The peak temperature obtained, also at axial location x=120 mm, using LPDF-EMST and LPDF-IEM models is close to 1400K which is approximately 7% lower than the experimental measurements. The LPDF-CD peak temperature, in this case, is close to 1320K. Even though the LPDF-CD predictions are significantly under-predicted in all three flames, we can observe an improvement in the predictions as the oxygen content in the coflow increases. As the oxygen content in the hot coflow increases from 3% (by mass) in HM1 to 6% in HM2 and to 9% in HM3 flames, the reaction rate improves due to which the models are able to predict the flames with better accuracy and a clear improvement in the temperature predictions, obtained using EDC and LPDF models, can be observed in the Figure 10. The under-predictions can directly be attributed to the lower reaction taking place along the centerline, in turn, over-predictions of $O_2$ profiles (Fig. 13).

Figures 11 and 12 depict the radial profiles of mean $CH_4$ and $H_2$ mass fraction obtained for HM1, HM2, and HM3 flames. As observed from the figures, EDC as well as LPDF models accurately capture the profiles, for all three flames, which can be attributed to well defined boundary conditions for fuel and co-flow streams. Major discrepancies can be observed in $O_2$ predictions for all three flames in Figure 13 that exhibits the radial profiles of mean $O_2$ mass fraction at different axial locations. The profiles are significantly over-predicted at the centerline in all three flames. Even after providing a better boundary conditions for oxygen (obtained from x=4 mm measurements) for all three flames, the evolution of $O_2$ could not be captured accurately which is evident from the over-estimation at the centerline, and thereby affecting the temperature predictions (Fig. 10). Along the centerline, the over-estimation can be attributed to the mixing between fuel jet and hot-coflow in the inner shear layer without sufficient reaction taking place. Whereas, the measurements in the jet region remains flat (~0) till downstream which essentially means complete reaction. This discrepancy is primarily due to the handling of reaction rate in modelling (turbulence-chemistry interaction) and cannot be quantified unless we look at the velocity statistics in the domain and this remains another drawback of this burner as it does not provide any velocity data. As we move from HM1 to HM3, we observe that this disagreement between the predictions and measurements increases. Among the LPDF models, there are no substantial differences observed between IEM and EMST predictions for all three flames, but as we move to HM2 and HM3, we observe the predictions appear to be in better shape, especially, for HM3 flame where the predictions, after over-predicting profiles at the centerline, capture the shear layer accurately as we move away radially. With an exception of HM2 flame, LPDF-CD predictions don't show substantial differences from the other two models as we move radially away from the centerline. This over-prediction of $O_2$ profiles at the centerline can be associated with the under-predictions of temperature at the centerline. Similar trends have already been reported in literature [12-14].

The radial profiles of mean $CO_2$ mass fraction obtained for HM1, HM2, and HM3 flames are shown in Figure 14. EDC as well as LPDF models adequately capture the profiles near the jet exit area (x=60 mm) in all three flames. EDC, LPDF-EMST, and LPDF-IEM capture the peak in profiles at the axial location x=60 mm in HM2 and HM3 flames whereas LPDF-CD fails to capture this peak. Discrepancies can be observed between the EDC and LPDF predictions away from the jet exit at x=120 mm location. The LPDF predictions agree with the measurements better than the EDC predictions at this location. EDC over-estimates the profiles in all three flames at this location (x=120 mm) which can be attributed to the over-estimation of mean reaction rate by EDC. The LPDF-EMST predictions are slightly over-predicted in HM3 at x=120 mm while LPDF-IEM and LPDF-CD accurately predict the profiles at this location. Radial profiles of mean $H_2O$ mass fraction for HM1, HM2, and HM3 flames are depicted in Figure 15 exhibiting a completely different picture from that of $CO_2$ profiles. EDC significantly under-predicts the profiles at centerline but captures them adequately as we move away from centerline radially for HM2 and HM3 flames; whereas for HM1 flames, the profiles remain under-predicted in the entire domain. As the oxygen content increases in the coflow from HM1 to HM3 flames, the extent of under-prediction at the centerline also increases. EDC under-predicts the profiles at centerline by approximately 30% at the axial location x=120 mm in HM1 flame whereas that value increases to approximately 36% in HM2 and 40% in HM3 at the same axial location. The LPDF-EMST and LPDF-IEM predictions show similar trend to that of EDC in all three flames with similar under-prediction at the centerline, although LPDF predictions are better than those of EDC. In case of LPDF-CD, the predictions for HM1 flame are similar to those of LPDF-EMST, and LPDF-IEM but as we move to HM2 and HM3 flames, the LPDF-CD predictions are significantly under-predicted in the shear layer. Mardani et al. [14] compared the EDC predictions obtained using DRM22 [22] and GRI 2.11 [23] chemical mechanisms with the experimental measurements for HM1 flame whereas Dongre et al. [18] compared MEPDF predictions obtained using DRM19 [22] and GRI 2.11 [23] chemical mechanisms against the measurements for HM1 flame. Both of these studies failed to observe any substantial differences between the predictions obtained using detailed chemical mechanism and the ones obtained using reduced mechanism. Thus, we can assert that chemical mechanism is not primarily responsible for the predictions obtained herein. However, it should be noted that the turbulent time scale is the dominant factor along the centerline which supersedes the scalar dissipation time scale and hence the combustible mixture dissipates quickly as the species are not allowed to stay for a longer duration in this region to complete reaction. This is one of the major reasons behind the discrepancies observed in species profiles along the centerline, especially the $O_2$ profiles and temperature (Figs 13, 10), in-turn, affecting the CO and $H_2O$ profiles that are under-predicted along the centerline. This slow chemistry increases the chemical time scales, in turn, reduces the Damkohler number and all the combustion models are unable to capture the flame characteristics in this low Damkohler number range which has significantly affected the species predictions.

To better understand the model behavior, it is worthwhile to look at the predictions of minor species like OH and CO. Major discrepancies can be observed in Figures 16 and 17 depicting OH and CO profiles, respectively. EDC fails to capture the CO and OH radicals. One particular observation here is that for OH predictions in HM1 and HM2 flames are significantly under-predicted by the EDC model near the jet exit (x=60 mm) but are over-predicted drastically in the downstream region (x=120 mm). EDC over-predicts OH profiles at x=120 mm by approximately 50% in HM1 flame which increases to 67% in HM2 and 75% in HM3 flames. This is, possibly, due to the entrainment of oxygen in the downstream region which enhances the fuel oxidation rate. None of the transported PDF models are able to accurately predict the CO and OH profiles for HM1 flame. Looking at the radial profiles of OH obtained for HM2 and HM3 flames, it can be seen here that the predictions are in better shape compared to those obtained for HM1 flame with LPDF models as the oxidation of $O_2$ into OH is sufficiently captured. The CO profiles are significantly under-predicted in the shear layer between fuel jet and hot coflow for HM2 and HM3 flames by EDC and LPDF models. The CO profiles are better in HM3 compared to those obtained in the HM2 flame owing to the better performance of LPDF combustion models due to improved reaction rates and oxygen contents in the coflow. In case of CO profiles, the LPDF predictions show substantial differences among themselves. With the exception of LPDF-CD predictions, other LPDF models are able to capture CO profiles adequately. Another observation is that the trend of CO predictions are consistent with measurements as we move from HM3 (9% $O_2$) to HM1 flame (3% $O_2$). As we move away from the centerline radially towards region where hot coflow and ambient cold coflow mix, we observe that EDC and LPDF models are capturing the CO profiles accurately. While comparing all of these finite rate based model, it has been observed that, in case of LPDF predictions, the differences between EMST and IEM predictions are small and better compared to the CD predictions which show considerable deviation from the two. A similar observation has been reported by Sarras et al. [44] for DJHC burner. From these Figures (13-17), it can be noted that the CO -> $CO_2$ conversion is significant resulting proper predictions of $CO_2$ and under-prediction of CO along the centerline; whereas OH -> $H_2O$ conversion is not significant resulting significant under-estimation of $H_2O$ along with the over-estimation of $O_2$ along the centerline. These phenomenon are strongly coupled between turbulence-chemistry interaction models and chemical mechanism which needs further investigation.

As done previously in the DJHC burner case, an extra scalar equation has been solved for EDC model using a user defined function to calculate temperature fluctuations and has been plotted against the RMS temperature profiles obtained using the LPDF models in Figure 18. The EDC model accurately captures the RMS temperature profiles for HM1 and HM2 flames but slightly over-predicts them at x=60mm location for HM3 flame. In case of the LPDF models, the EMST and IEM predictions are in a better shape than the CD predictions although all three LPDF models under-predict the peak RMS temperature in the downstream region. Unlike the DJHC case, the inlet boundary conditions of RMS temperature do not exhibit significant effect on the predictions for this burner. The radial profiles of RMS $CO_2$ mass fraction obtained for HM1, HM2, and HM3 flames are depicted in Figure 19. Even though the LPDF models slightly under-predict the RMS profiles at the centerline near the jet exit region (x=60 mm), the profiles are adequately captured away from the centerline at x=60 mm for HM1 and HM2 flames. The peaks of the profiles for HM1, HM2, and HM3 flames are shifted radially compared to measurements which can be attributed to errors in capturing mixing between the fuel and coflow stream in the shear layer. As we move away from the jet exit area axially in the downstream region (x=120 mm), the extent of under-prediction at the centerline reduces as oxygen content in the coflow increases (a trend which is also observed at the axial location x=60 mm near the jet exit in all three flames). The profile peaks at axial location x=120 mm are significantly under-predicted in all three flames. In HM1 flame, the LPDF models under-predict the profiles by approximately 30% whereas in HM2 and HM3 flames, it is approximately 27% and 24%, respectively, thus, showcasing the effect of improved oxygen content of the coflow on the predictions. A similar trend of results, as seen in $CO_2$ RMS profiles, is observed in $H_2O$ RMS profiles as shown in Figure 20. The shifted peaks near the jet exit area (x=60 mm) and under-prediction of the peak profiles in the downstream region (x= 120 mm) are observed here. The main differences between the RMS $CO_2$ and $H_2O$ profiles are observed at x=120 mm location. In case of RMS $CO_2$ mass fraction profiles, the LPDF predictions, with different micro-mixing models, show no substantial differences among them and under-predict the profiles at the centerline. In case of RMS $H_2O$ mass fraction profiles, the LPDF predictions don't show substantial differences among themselves and, with exception of LPDF-CD, other LPDF models accurately capture the profiles along the centerline.

The radial profiles of RMS $O_2$ mass fraction obtained for HM1, HM2, and HM3 flames are shown in Figure 21. Comparing the RMS profiles with the mean profiles depicted in Figure 13, we observe that the RMS profiles are accurately predicted at the centerline, at both locations i.e. near the jet exit (x=60 mm) and in the downstream region (x=120 mm), unlike the mean profiles. The RMS profiles are accurately captured near the jet exit area at axial location x=60 mm with profile peaks slightly shifted radially because of inaccuracies in predicting mixing between the fuel and coflow streams in the shear layer; whereas the peaks are slightly under-predicted at the downstream region (x=120 mm). Figures 22 and 23 showcase the radial RMS profiles of CO and OH mass fraction obtained for HM1, HM2, and HM3 flames. None of the LPDF models are able to predict the RMS profiles of CO and OH accurately in the entire domain for HM1 flame. Predictions improve for HM2 and HM3 flames where LPDF models are able to adequately capture the trends of the measurements. For both the CO and OH RMS profiles, the LPDF-CD predictions are significantly under-predicted as compared to IEM and EMST predictions for HM2 flame at both x=60 and 120 mm locations. In case of HM3 flame, CD predictions are significantly under-predicted at x=60 mm location whereas at x=120 mm location, we observe a sharp rise in CD predictions although they are still under-predicted and are on the lower side of IEM and EMST predictions. This sharp rise in CD predictions in the downstream location (x=120 mm) can be due to increased entrainment of oxygen from the ambient cold coflow and affecting the predictions due to the nature of this mixing model as discussed below.

The discrepancies observed, so far, with the finite rate chemistry based models arise due to multiple reasons. Slower reaction rates in MILD combustion, compared to standard combustion processes, make it more challenging for modelling. EDC predictions are nice on the fuel lean side but over predict on the fuel rich side. This is due to the fact that the EDC model with default time scale constant does not capture the slower reaction rate properly, especially in MILD combustion regime [5]. In turn, an early ignition takes place at the radial location where turbulence is low and therefore the effects of turbulence-chemistry get undermined in this region as a result of which it shifts the profiles upstream and towards the axis. Despite having the chemical source terms in closed form, the transported PDF models cannot also be precisely accurate as the major source of errors in these models comes from the inaccuracies of the micro-mixing closures. The predictions in the shear layer are primarily affected due to mixing between the hot coflow and the fuel jet. This mixing between the streams, due to turbulence and species gradients, emphasizes the role of micro-mixing here. In case of IEM model [33], the composition of all the scalars relaxes towards the mean composition at the same rate whereas in case of CD mixing model, a number of particles in a cell are randomly selected and their individual compositions are moved towards the mean composition [34]. Therefore, both of these mixing models are missing the effect of localness in the flow field which is a potential source of errors for the discrepancies observed in the predictions obtained through these models. However, on the contrary, the EMST model [21] takes into account the effects of local mixing among the particles in the composition space and thereby making it more accurate compared to the CD and the IEM models. The differences among the mixing models can be further analyzed by studying scatter plots of temperature and species mass fractions against the mixture fraction which are depicted in Figures 24 and 25, exhibit scatter plots for temperature, $CO_2$, CO, $O_2$, OH, and $H_2O$ mass fractions in the domain for all three flames. From the figures, it is clearly visible that the variance is low with the EMST model and it further reduces as we move from HM1 flame towards HM3 flame. This can be attributed to the localness of the EMST model in the composition space as explained above. Even though the variance with the IEM model is more compared to the EMST model due to non-localness in the composition space, it's magnitude of variance in the vertical direction is similar to that of the EMST model for temperature as well as species mass fractions which quantifies the mean predictions observed in Figure 10 for temperature and Figures 13-17 for $O_2$, $CO_2$, $H_2O$, OH, and CO mass fractions. Since, in case of CD mixing model, the particles, in a cell, are selected randomly, CD formulation is missing the effects of localness in the composition space which is the main reason behind the differences observed between EMST and CD predictions and is, also, evident from the scatter plots. Another potential source of errors may arise from the 'notional particles', used in Lagrangian PDF approach: one due to particle tracking scheme, and another one due to Monte-Carlo methods. The mean density of particles in physical space should remain proportional to the local mean fluid density all the times and this can be satisfied as long as the particle systems evolve in consistent manner with the Eulerian equation systems. Therefore, the accuracy of the particle tracking scheme may induce some errors. Second source of discrepancies in the LPDF predictions is the numerical errors associated with the Monte-Carlo methods, e.g. statistical errors, bias errors. Statistical errors are the random errors whereas bias errors are deterministic errors and both of them are strongly dependent on the number of particles used per cell. Statistical errors are not present in the present simulations (Figs 7-8); however, the bias errors, which arise from the mean quantities, are not completely reduced by averaging [32].

In the MILD combustion regime, the initiation of reaction is delayed and overall reaction rates are lower than the conventional flames with lower temperatures and $NO_x$ emissions. For the HM1 flame, there's 3% oxygen in the coflow, which is too much dilution for a JHC flame and hence the HM1 flame forms a very crude case for studying the combustion models in the MILD regime. Thus, comparing all the flames spread over wide range of $O_2$% (3-9) at the end, it is clearly evident that predictions improve as the oxygen content of the coflow increases, the reaction rate also increases which, in-turn, increases the Damkohler number. Therefore, we can say that we obtain better results at high Damkohler number irrespective of the chosen turbulence-chemistry interaction models. The EDC model slightly over-predicts the temperature profiles, although it properly captures the mean mixture fraction and major species. The transported PDF models reasonably capture the RMS profiles in the entire domain with the best results obtained for HM3 flames owing to better oxygen content in the coflow. Overall, the mean predictions obtained through transported PDF based models are in good agreement with the measurements whereas the RMS profiles have been adequately captured with discrepancies observed mostly in the shear layer between the fuel jet and the hot coflow.

## 5. CONCLUSIONS

In the present study, two different burners imitating MILD combustion have been investigated using SF, EDC, LPDF, and MEPDF turbulence-chemistry interaction models. In the first part of the study, DJHC-I flame corresponding to Reynolds number Re=4100 has been simulated. Steady flamelet is not able to predict the scalar field correctly due to single-mixture fraction formulation whereas EDC model over-predicts the reaction rate resulting in discrepancies in the downstream region. The LPDF models accurately capture the mean as well as RMS profiles. In context of the MEPDF model, the predictions obtained for DJHC-I flame (Re=4100) are in good agreement with the experimental measurements with an over prediction of temperature in the downstream region. It has been observed that, in case of transported PDF models, it is not sufficient to provide only the mean temperature as boundary conditions but temperature fluctuations need to be also included in order to obtain better predictions. .

In the later part of the study, three Adelaide JHC flames, namely HM1, HM2 and HM3, all corresponding to Reynolds number Re=10000, have been simulated. The mean mass fraction profiles of major species like $CH_4$, $CO_2$, $H_2$, $H_2O$, and $O_2$ along with those of radicals like CO and OH along with profiles of mean temperature are reported at different axial and radial locations. Both SF and MEPDF models fail to capture the major as well as minor species at different axial locations. The predictions obtained using the EDC and LPDF models are in good agreement with the experimental measurements with minor discrepancies observed in predictions of mean temperature and species like $CO_2$, $H_2O$, and $O_2$. Major discrepancies have been observed while predicting minor species like CO and OH. LPDF models adequately capture the RMS profiles of temperature, $CO_2$, $H_2O$, and $O_2$ mass fractions whereas the poor predictions of CO and OH RMS profiles improve with improving oxygen content of the coflow. In case of the LPDF predictions, the EMST and IEM predictions are found closer to each other and in better agreement with the measurements than the CD predictions which are found to be inferior to the other models. Based on the results obtained, we can conclude that the transported PDF models are also not free from errors and still provide the best results for JHC flames when compared with other models, while the results obtained with the EDC model can be improved by increasing the chemical time scale constant, which increases the delay in initiation of the reaction, a desirable aspect in the MILD combustion.

## 6. ACKNOWLEDGEMENTS

Simulations are carried out on the computers provided by the Indian Institute of Technology Kanpur (IITK) (www.iitk.ac.in/cc) and the manuscript preparation as well as data analysis has been carried out using the resources available at IITK. This support is gratefully acknowledged. The authors would also like to thank Prof. Bassam Dally of the Adelaide University for providing the experimental database for the Adelaide burner.

**List of figures**

Fig. 1 Radial profiles of mean axial velocity, mean temperature, and turbulent kinetic energy for grid independence study of DJHC burner obtained using EDC model. Symbols (*triangle* $0 \leq r \leq 35$, *squares* $-35 \leq r \leq 0$) are measurements and lines are predictions.

Fig. 2 Radial profiles of mean axial velocity, mean temperature and turbulent kinetic energy obtained for DJHC-I flame (Re=4100) with $C_\varphi =2$. Symbols (*triangle* $0 \leq r \leq 35$, *squares* $-35 \leq r \leq 0$) are measurements and lines are predictions.

Fig. 3 Radial profiles of Reynolds stresses obtained for DJHC-I flame (Re=4100) with $C_\varphi=2$. Symbols (*triangle* $0 \leq r \leq 35$, *squares* $-35 \leq r \leq 0$) are measurements and lines are predictions.

Fig. 4 Radial RMS profiles of temperature obtained for DJHC-I flame (Re=4100) with $C_\varphi=2$. Symbols (*triangle* $0 \leq r \leq 35$, *squares* $-35 \leq r \leq 0$) are measurements and lines are predictions.

Fig. 5 Radial and centerline profiles of RMS mean mixture fraction for grid independence study for HM1 flame at Re=10000. Symbols (*triangle* $0 \leq r \leq 70$, *squares* $-25 \leq r \leq 0$) are measurements and lines are predictions.

Fig. 6 Radial profiles of $CO_2$, $H_2O$, OH, and CO mass fraction for HM1 flame with $C_\varphi =2$. Symbols (*triangle* $0 \leq r \leq 70$, *squares* $-25 \leq r \leq 0$) are measurements and lines are predictions. The OH and CO mass fraction profiles have been scaled as Y(OH)/8 and Y(CO)/5.

Fig. 7 Radial profiles of $H_2O$ and $O_2$ mean mass fraction for HM1 flame obtained using LPDF-EMST model. Symbols (*triangle* $0 \leq r \leq 70$, *squares* $-25 \leq r \leq 0$) are measurements and lines are predictions.

Fig. 8 Radial RMS profiles of H$_2$O and O$_2$ mass fraction for HM1 flame obtained using LPDF-EMST model. Symbols (*triangle* $0 \leq r \leq 70$, *squares* $-25 \leq r \leq 0$) are measurements and lines are predictions.

Fig. 9 Radial profiles of mean mixture fraction for (a) HM1, (b) HM2, and (c) HM3 flames with $C_\varphi$ =2. Symbols (*triangle* $0 \leq r \leq 70$, *squares* $-25 \leq r \leq 0$) are measurements and lines are predictions.

Fig. 10 Radial profiles of mean temperature for (a) HM1, (b) HM2, and (c) HM3 flames with $C_\varphi$ =2. Symbols (*triangle* $0 \leq r \leq 70$, *squares* $-25 \leq r \leq 0$) are measurements and lines are predictions.

Fig. 11 Radial profiles of CH$_4$ mass fraction for (a) HM1, (b) HM2, and (c) HM3 flames with $C_\varphi$ =2. Symbols (*triangle* $0 \leq r \leq 70$, *squares* $-25 \leq r \leq 0$) are measurements and lines are predictions.

Fig. 12 Radial profiles of H$_2$ mass fraction for (a) HM1, (b) HM2, and (c) HM3 flames with $C_\varphi$ =2. Symbols (*triangle* $0 \leq r \leq 70$, *squares* $-25 \leq r \leq 0$) are measurements and lines are predictions.

Fig. 13 Radial profiles of O$_2$ mass fraction for (a) HM1, (b) HM2, and (c) HM3 flames with $C_\varphi$ =2. Symbols (*triangle* $0 \leq r \leq 70$, *squares* $-25 \leq r \leq 0$) are measurements and lines are predictions.

Fig. 14 Radial profiles of CO$_2$ mass fraction for (a) HM1, (b) HM2, and (c) HM3 flames with $C_\varphi$ =2. Symbols (*triangle* $0 \leq r \leq 70$, *squares* $-25 \leq r \leq 0$) are measurements and lines are predictions.

Fig. 15 Radial profiles of H$_2$O mass fraction for (a) HM1, (b) HM2, and (c) HM3 flames with $C_\varphi$

=2. Symbols (*triangle* $0 \leq r \leq 70$, *squares* $-25 \leq r \leq 0$) are measurements and lines are predictions.

Fig. 16 Radial profiles of OH mass fraction for (a) HM1, (b) HM2, and (c) HM3 flames with $C_\varphi$ =2. Symbols (*triangle* $0 \leq r \leq 70$, *squares* $-25 \leq r \leq 0$) are measurements and lines are predictions.

Fig. 17 Radial profiles of CO mass fraction for (a) HM1, (b) HM2, and (c) HM3 flames with $C_\varphi$ =2. Symbols (*triangle* $0 \leq r \leq 70$, *squares* $-25 \leq r \leq 0$) are measurements and lines are predictions.

Fig. 18 Radial RMS profiles of temperature for (a) HM1, (b) HM2, and (c) HM3 flames with $C_\varphi$ =2. Symbols (*triangle* $0 \leq r \leq 70$, *squares* $-25 \leq r \leq 0$) are measurements and lines are predictions.

Fig. 19 Radial RMS profiles of $CO_2$ mass fraction for (a) HM1, (b) HM2, and (c) HM3 flames with $C_\varphi$ =2. Symbols (*triangle* $0 \leq r \leq 70$, *squares* $-25 \leq r \leq 0$) are measurements and lines are predictions.

Fig. 20 Radial RMS profiles of $H_2O$ mass fraction for (a) HM1, (b) HM2, and (c) HM3 flames with $C_\varphi$ =2. Symbols (*triangle* $0 \leq r \leq 70$, *squares* $-25 \leq r \leq 0$) are measurements and lines are predictions.

Fig. 21 Radial RMS profiles of $O_2$ mass fraction for (a) HM1, (b) HM2, and (c) HM3 flames with $C_\varphi$ =2. Symbols (*triangle* $0 \leq r \leq 70$, *squares* $-25 \leq r \leq 0$) are measurements and lines are predictions.

Fig. 22 Radial RMS profiles of CO mass fraction for (a) HM1, (b) HM2, and (c) HM3 flames with $C_\varphi$ =2. Symbols (*triangle* $0 \leq r \leq 70$, *squares* $-25 \leq r \leq 0$) are measurements and lines are predictions.

Fig. 23 Radial RMS profiles of OH mass fraction for (a) HM1, (b) HM2, and (c) HM3 flames with $C_\varphi$ =2. Symbols (*triangle* $0 \leq r \leq 70$, *squares* $-25 \leq r \leq 0$) are measurements and lines are predictions.

Fig. 24 Scatter plots of temperature, $CO_2$, CO mass fractions obtained for (a) HM1, (b) HM2, and (c) HM3 flames with $C_\varphi$ =2.

Fig. 25 Scatter plots of $O_2$, OH, and $H_2O$ mass fractions obtained for (a) HM1, (b) HM2, and (c) HM3 flames with $C_\varphi$ =2.

# Figures

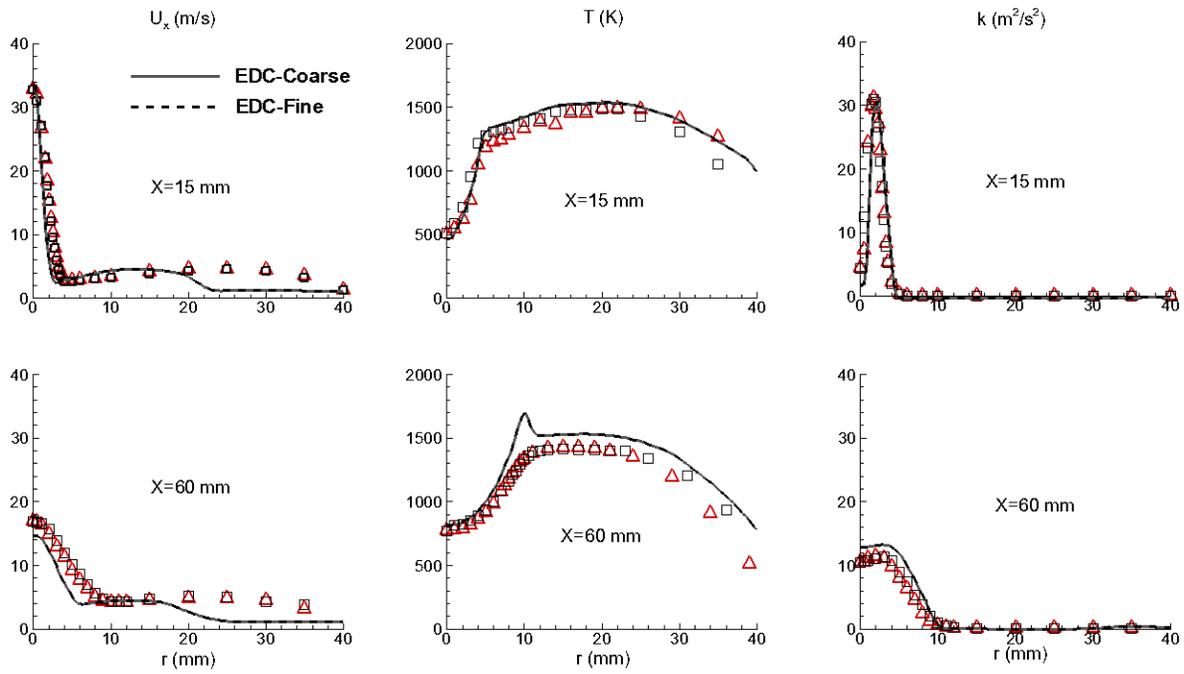

Figure 1

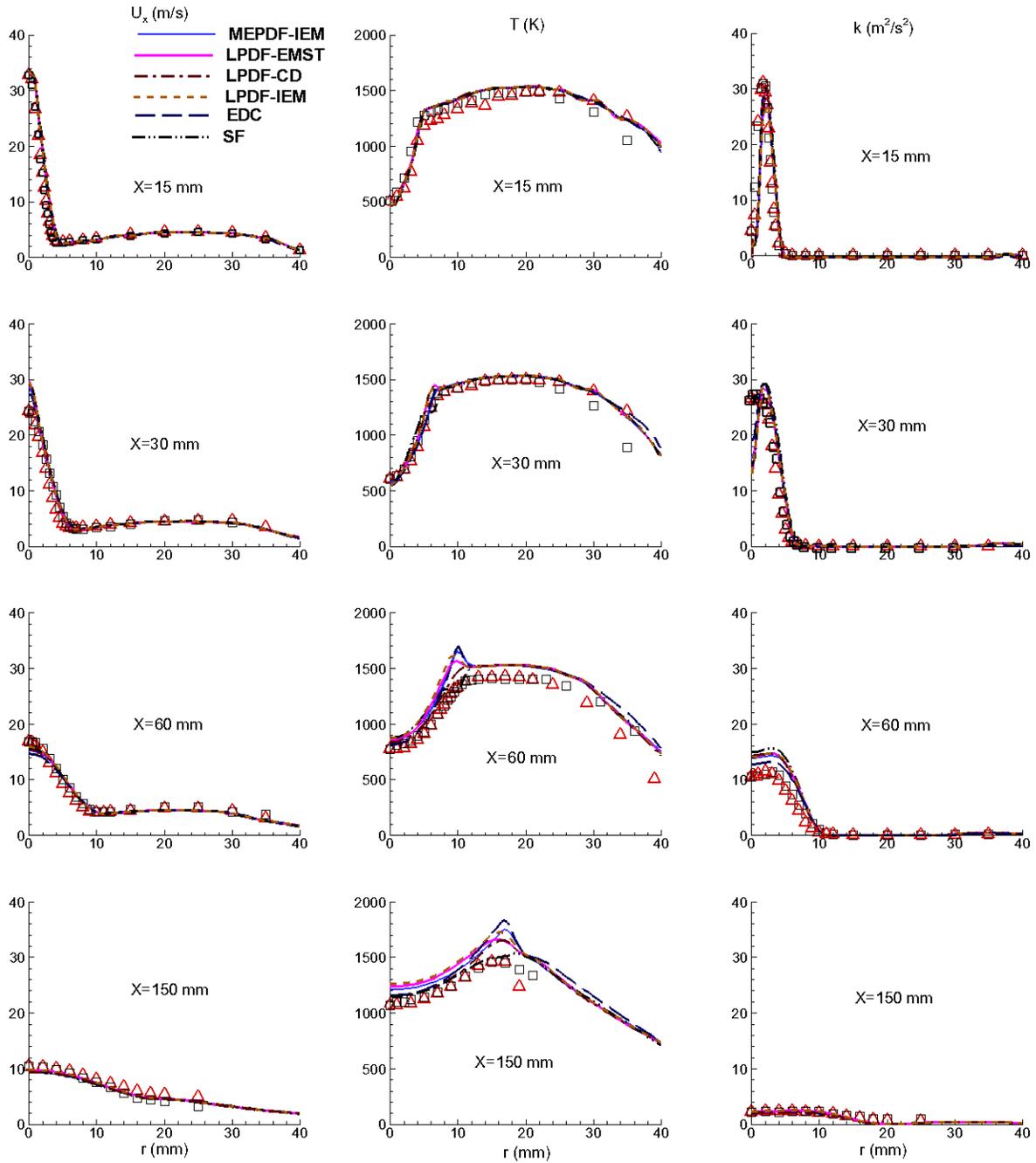

Figure 2

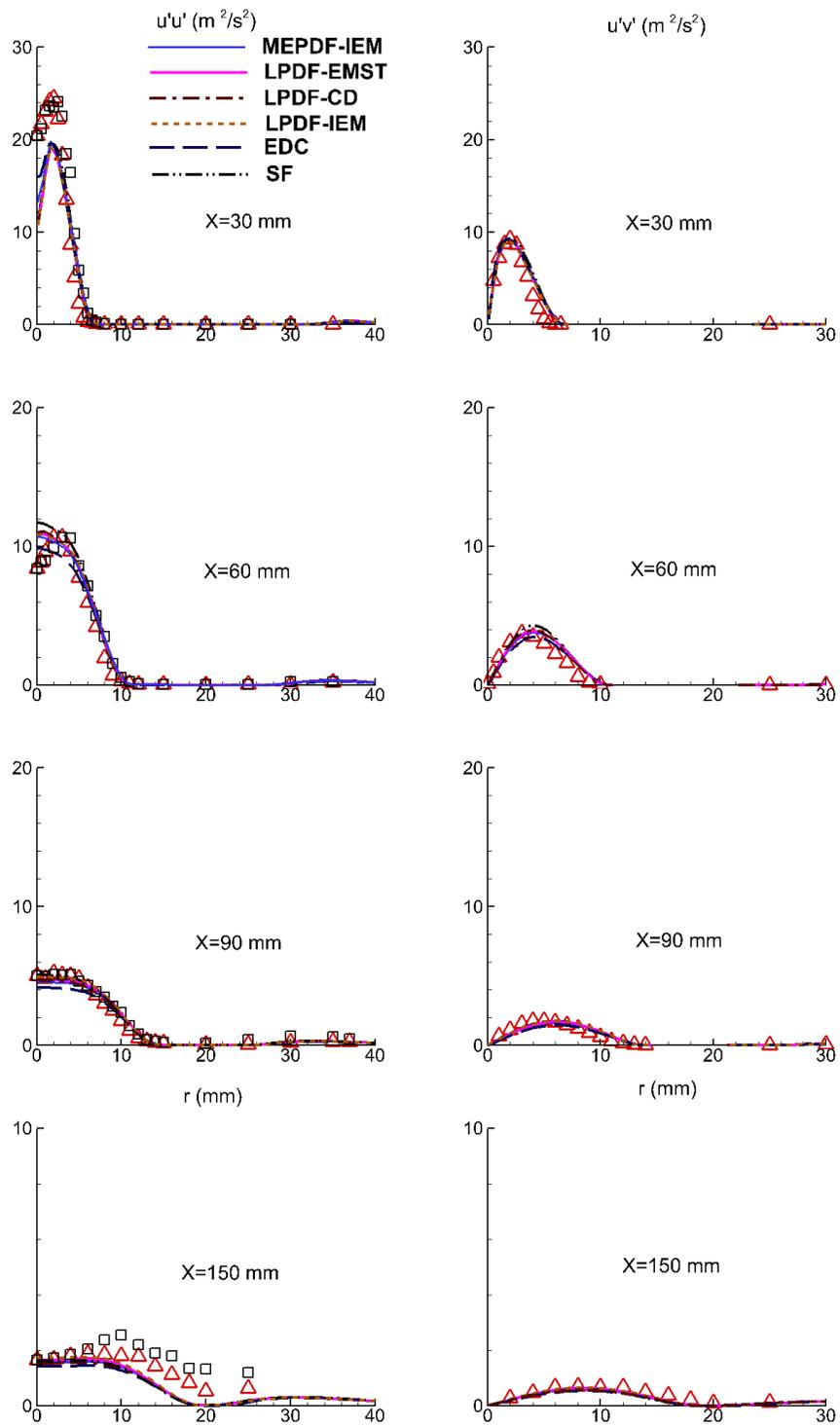

Figure 3

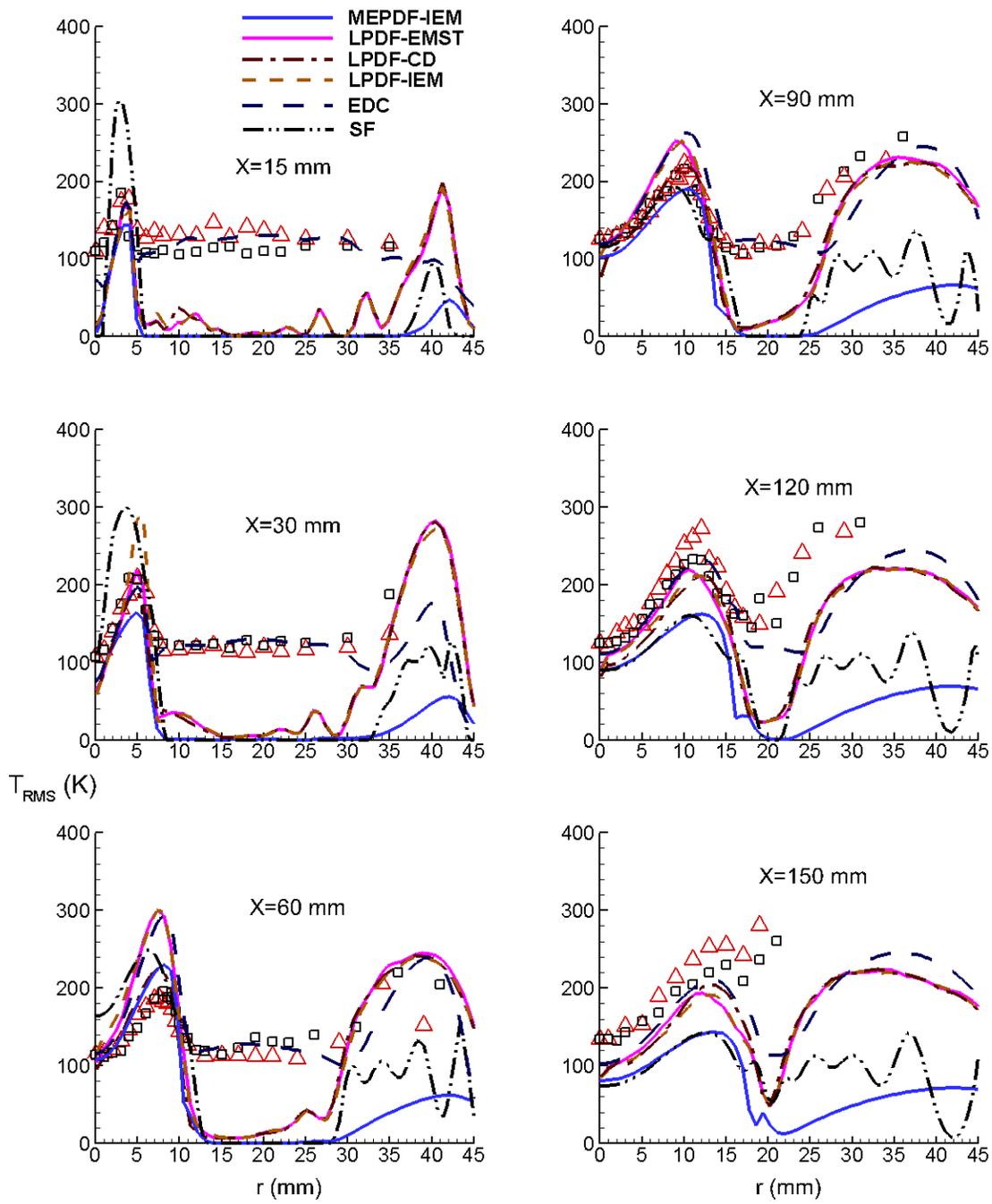

Figure 4

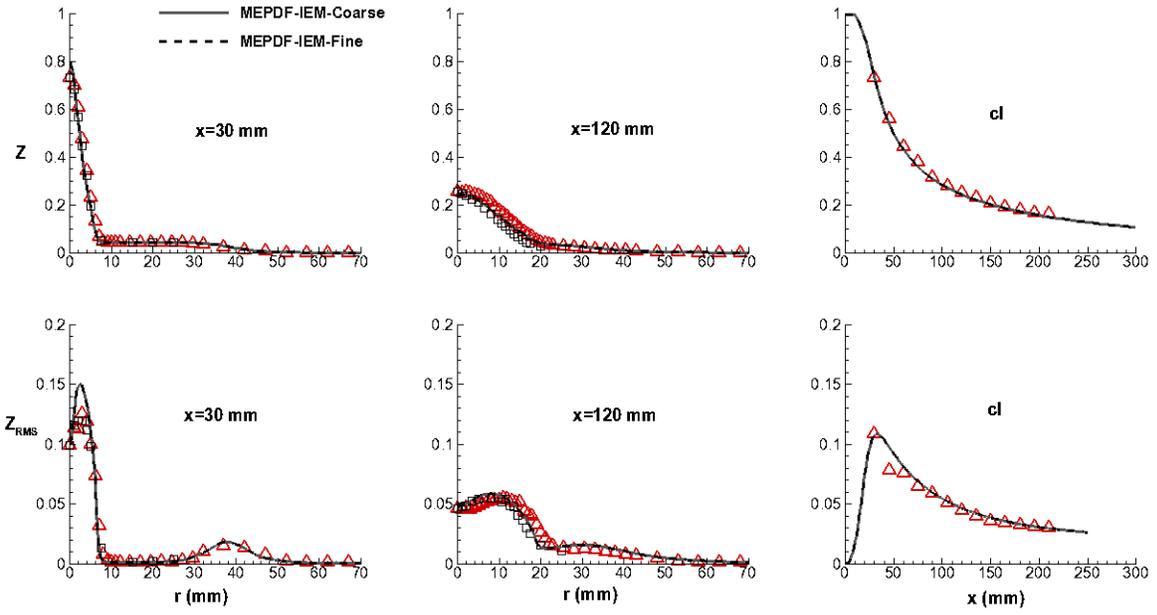

Figure 5

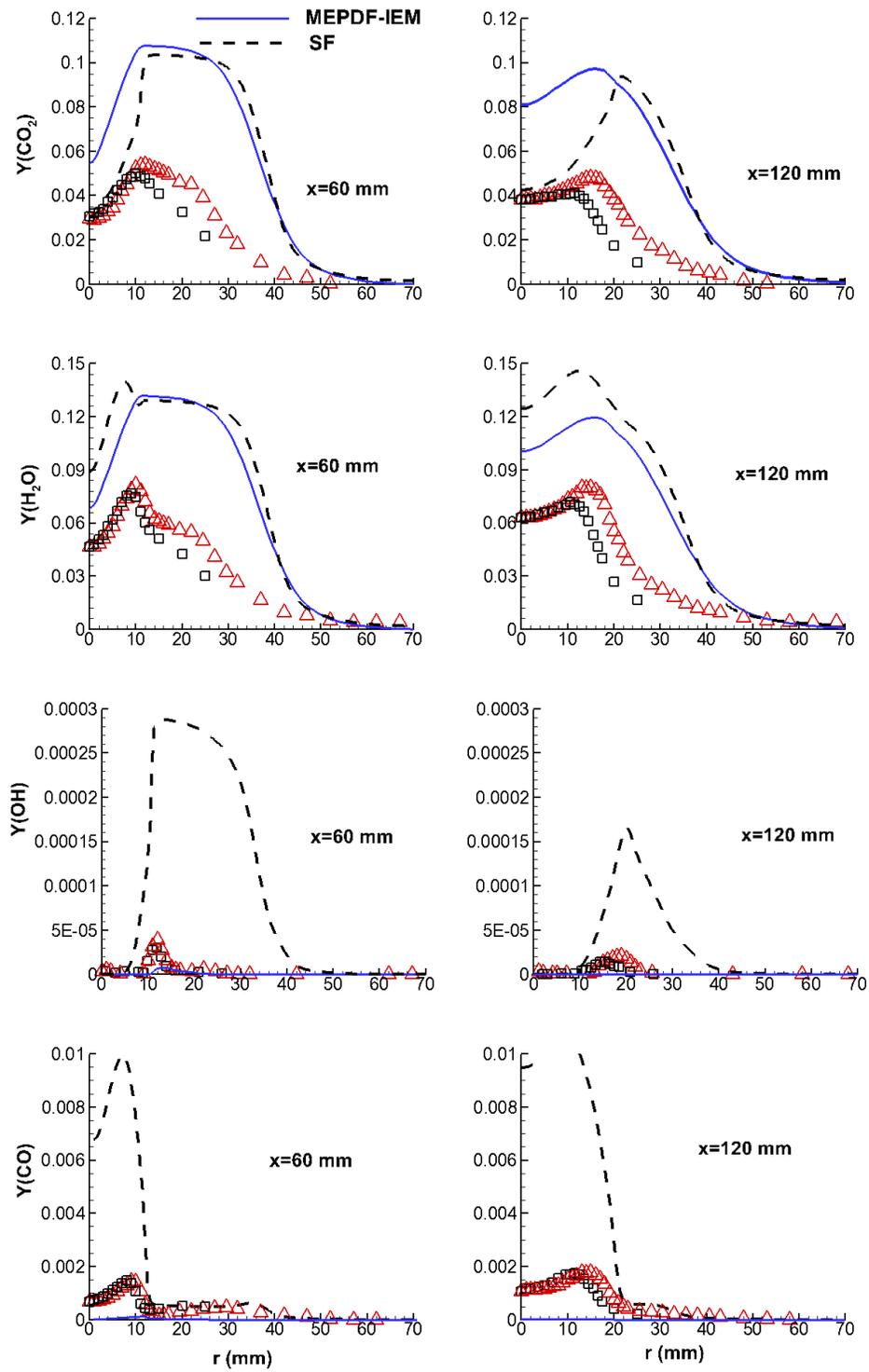

Figure 6

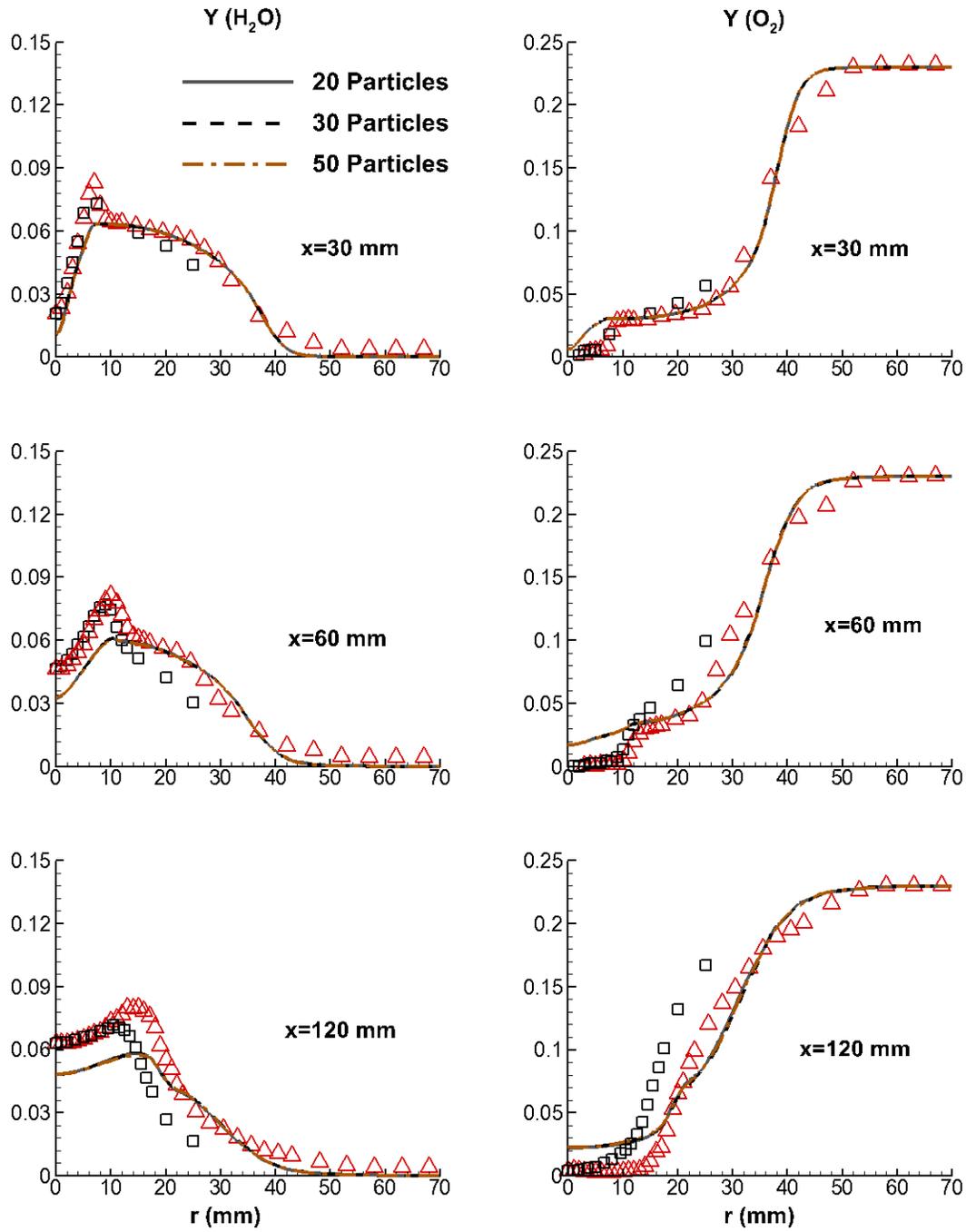

Figure 7

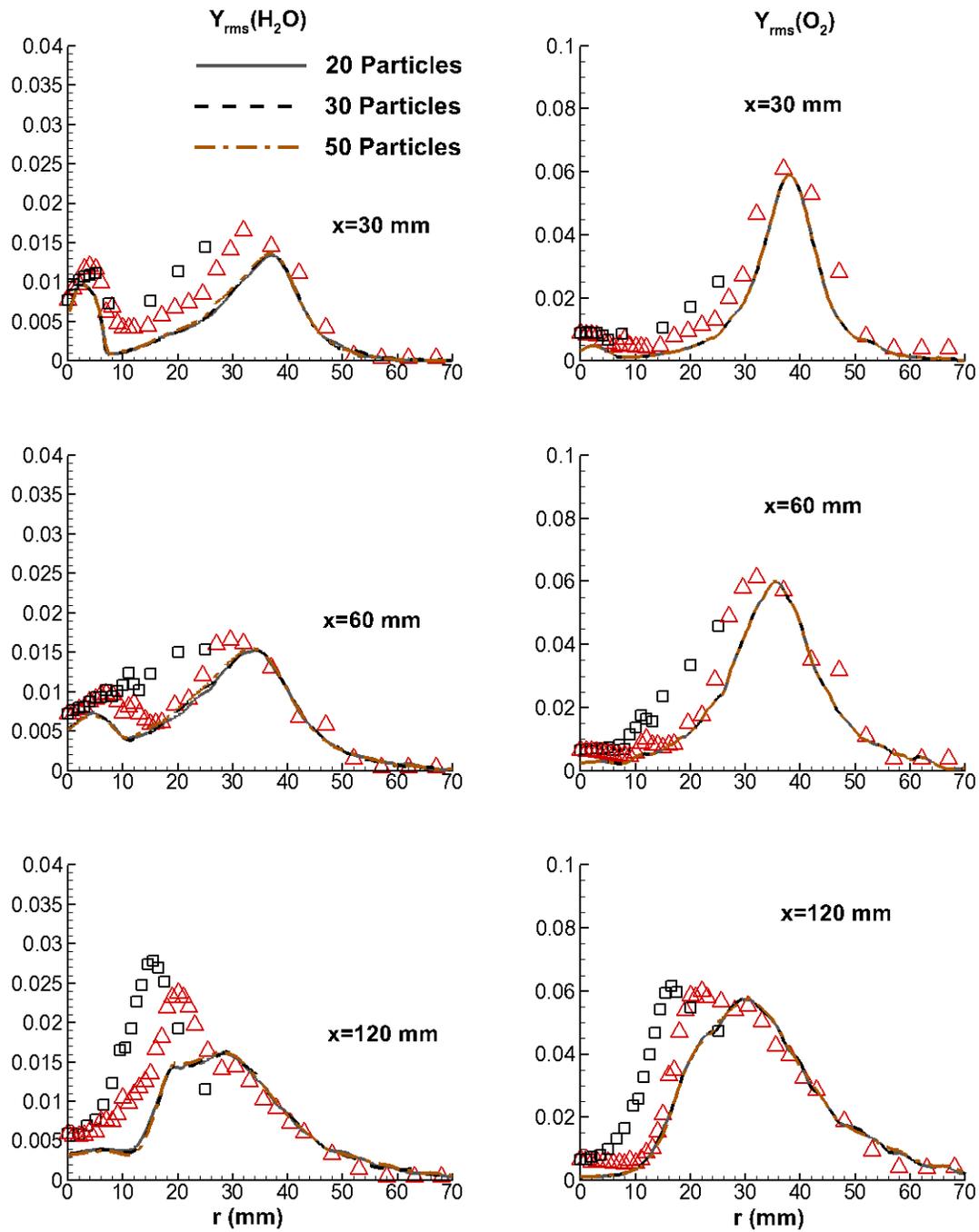

Figure 8

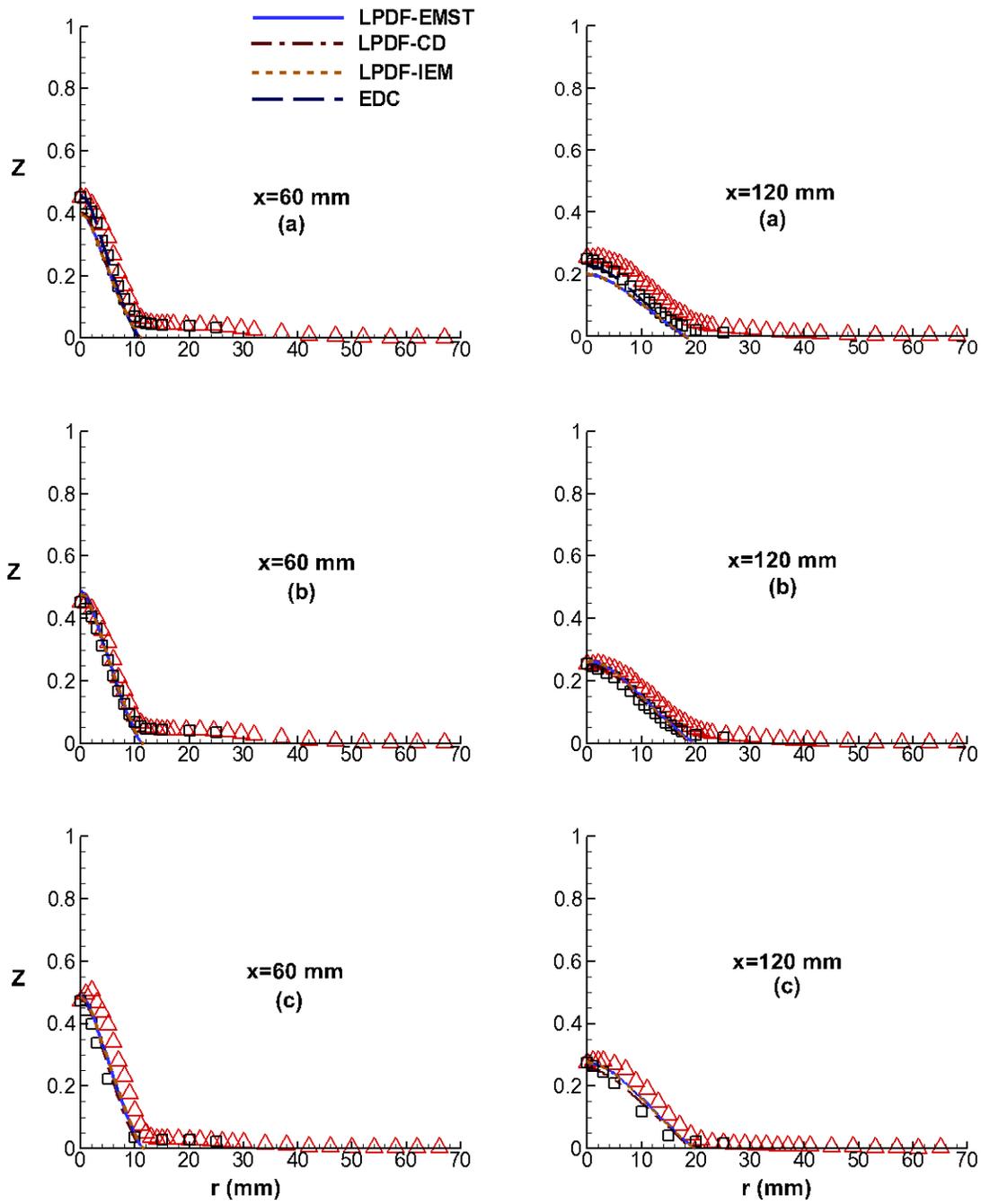

Figure 9

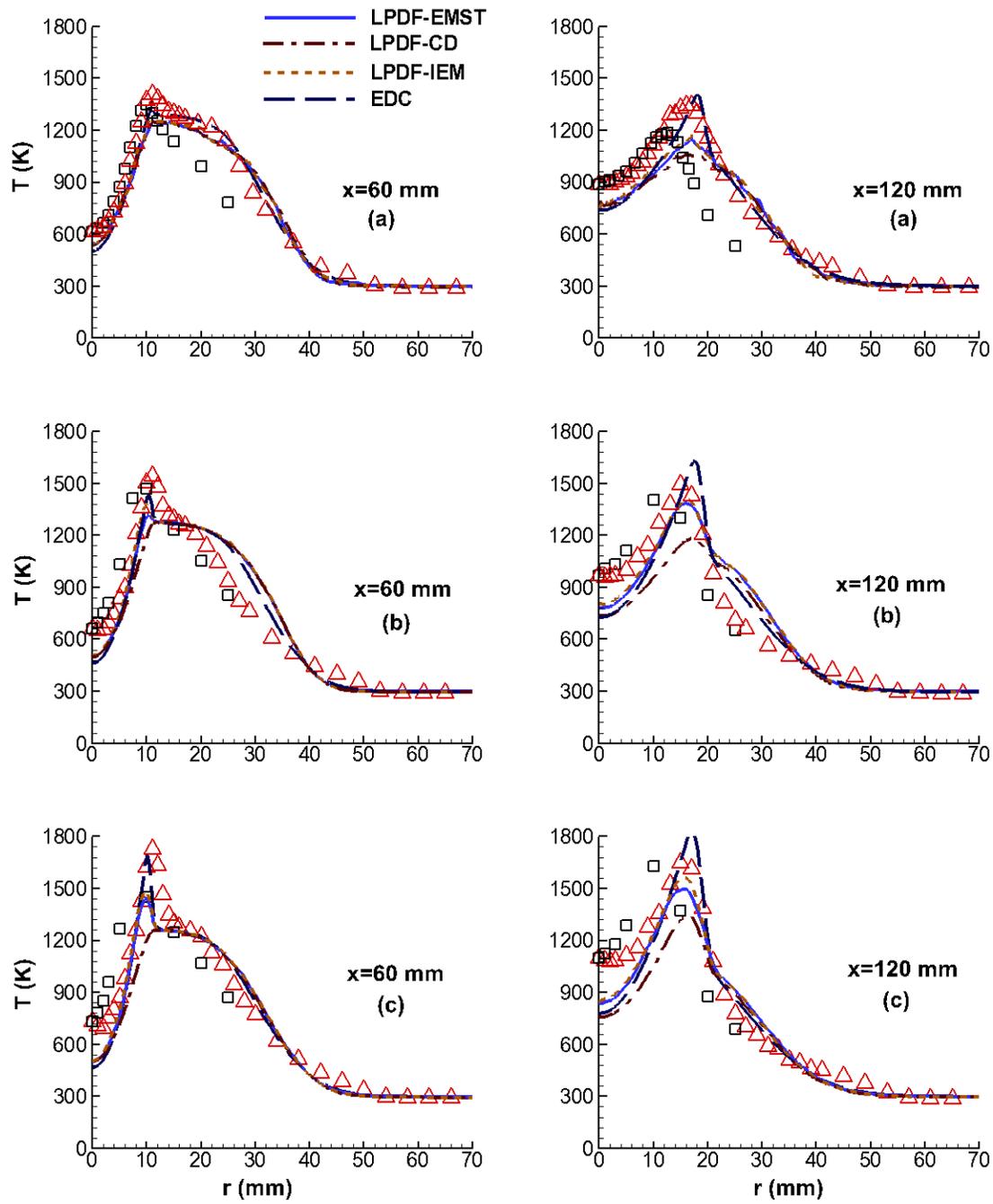

Figure 10

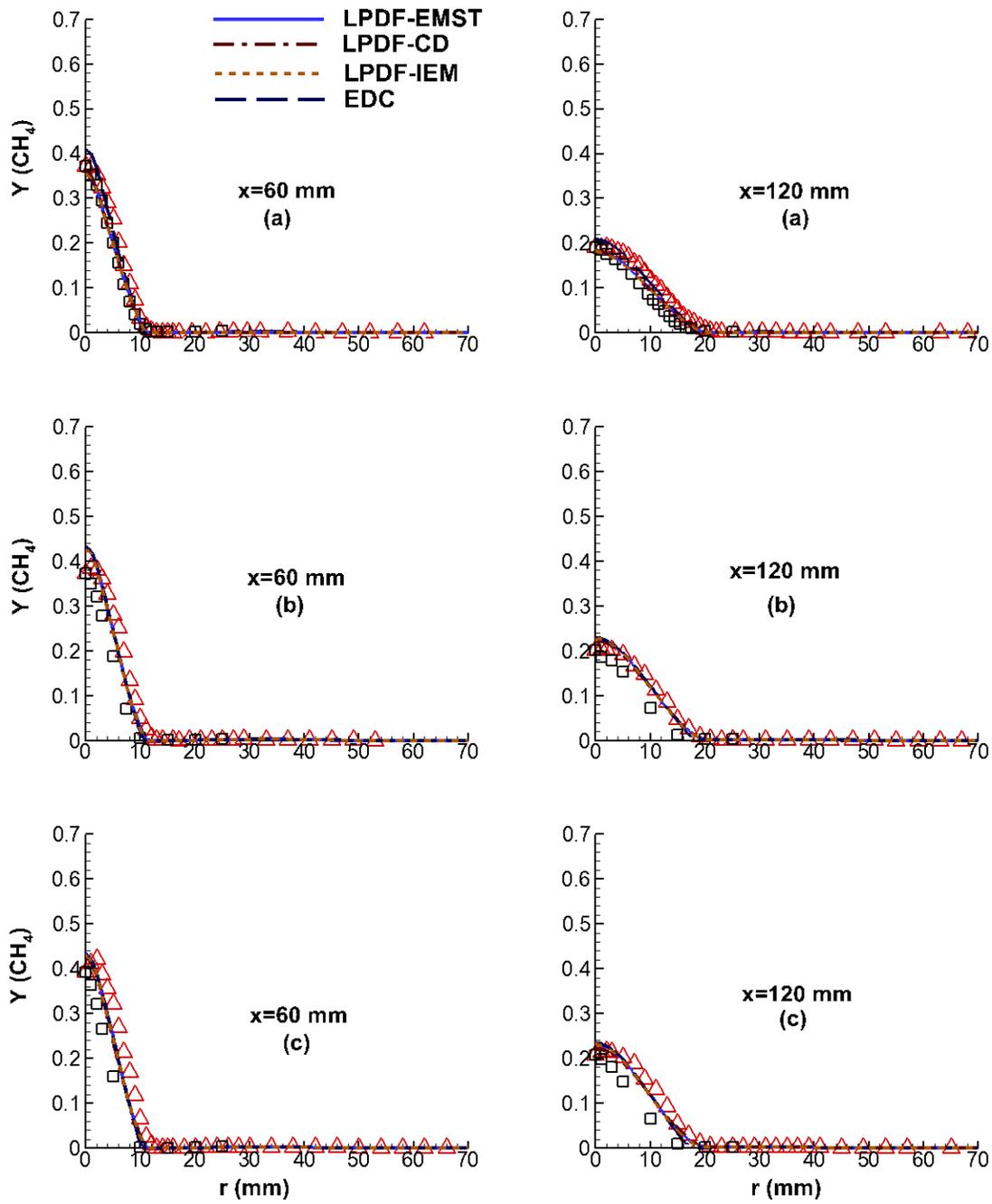

Figure 11

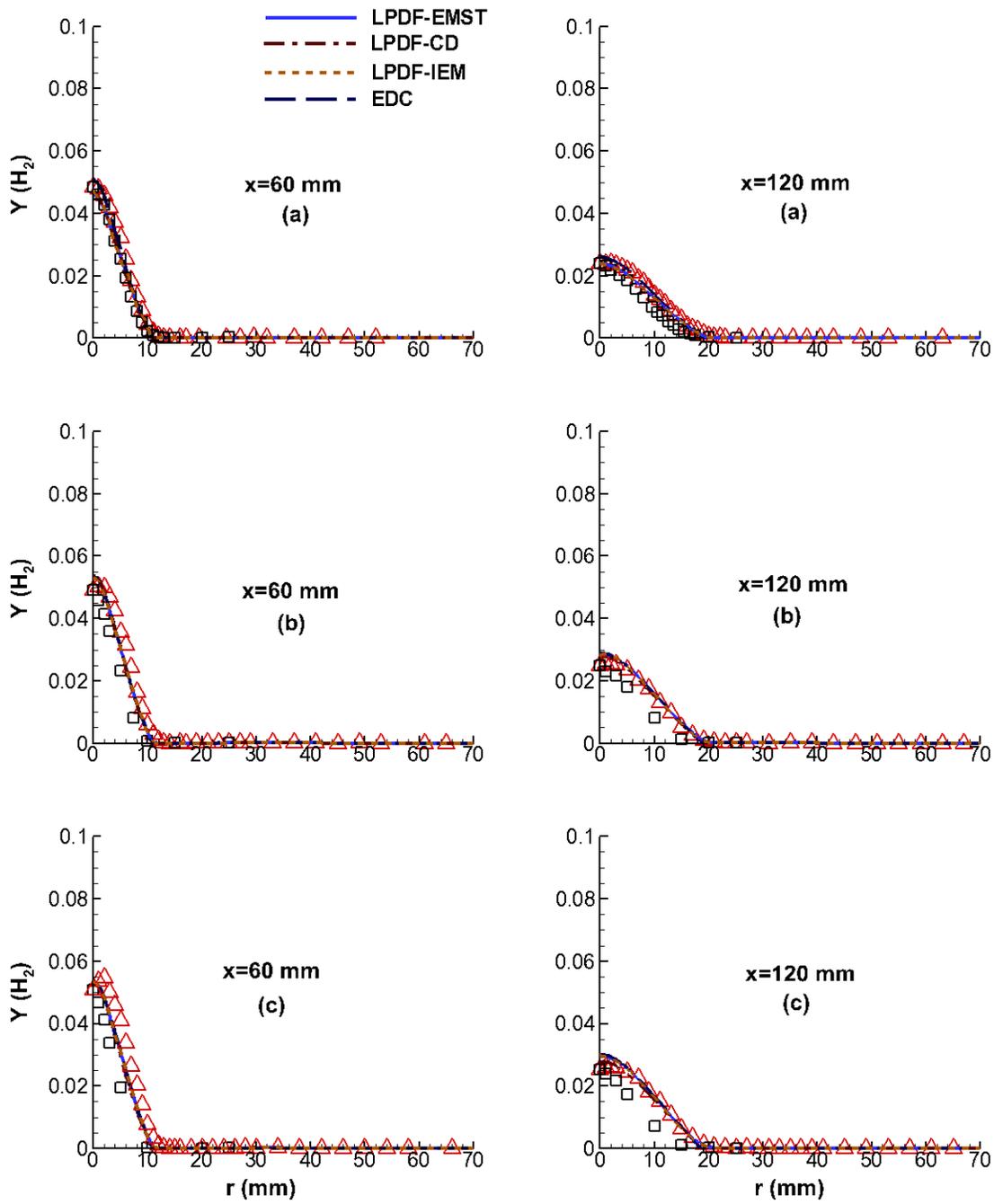

Figure 12

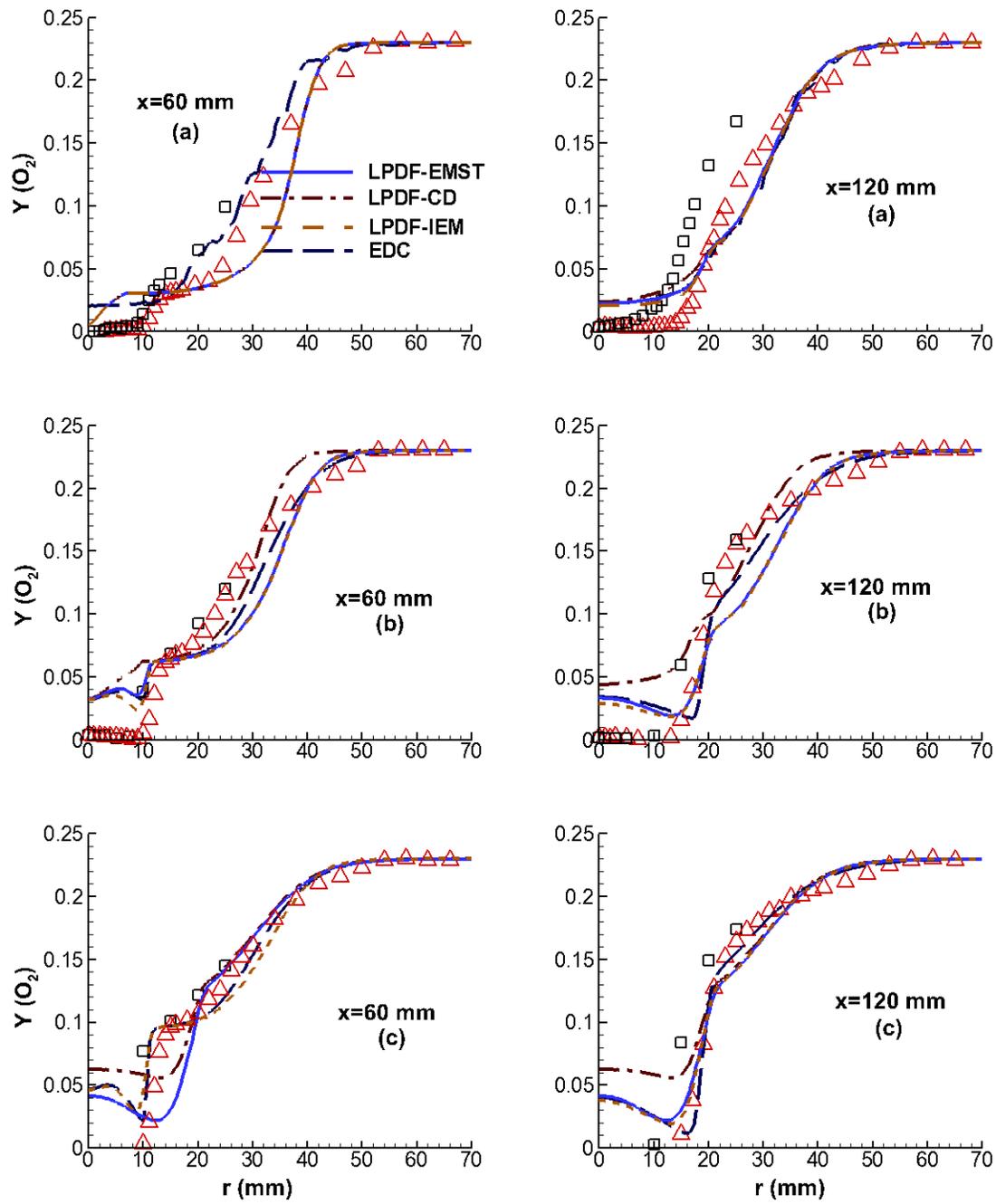

Figure 13

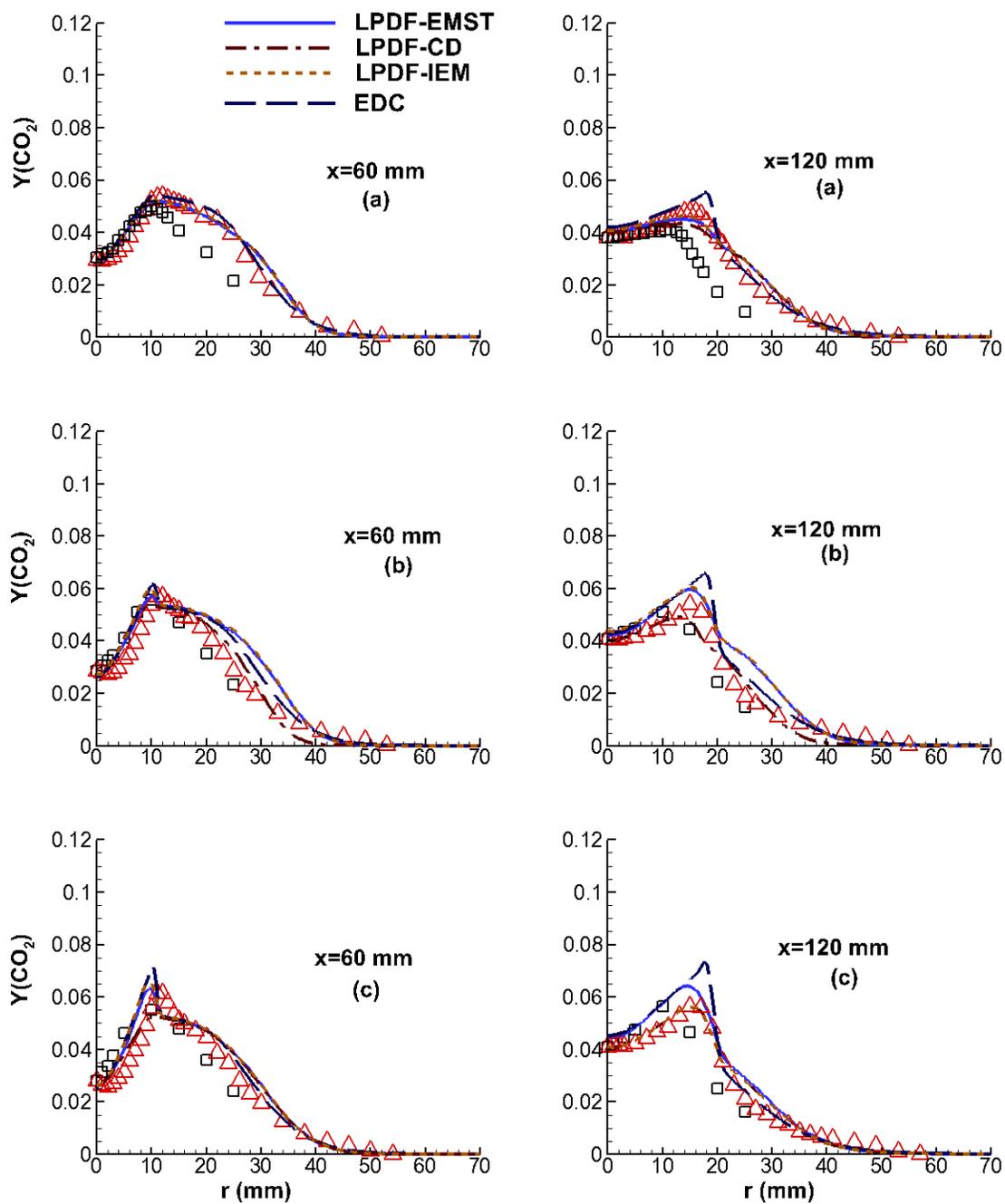

Figure 14

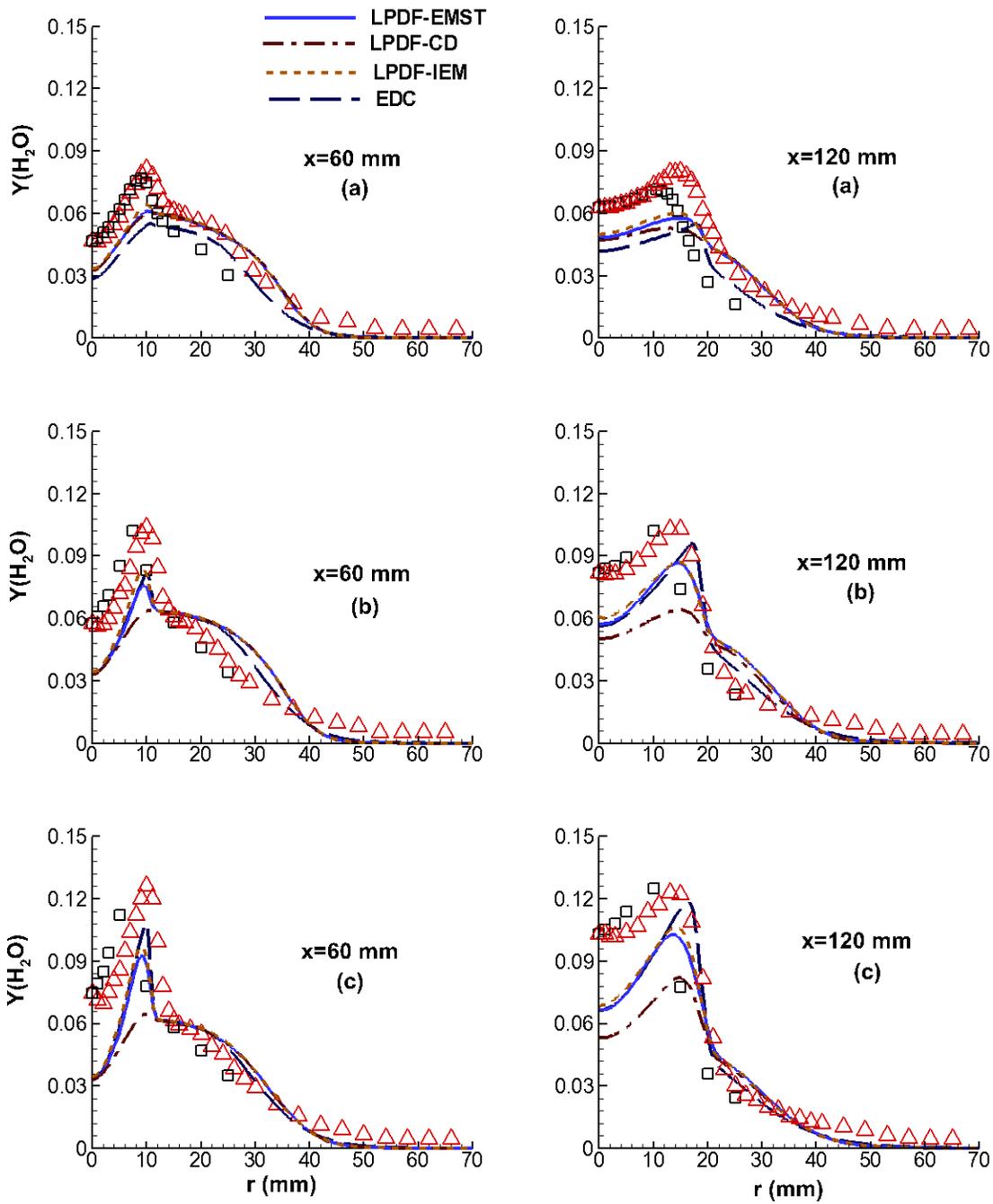

Figure 15

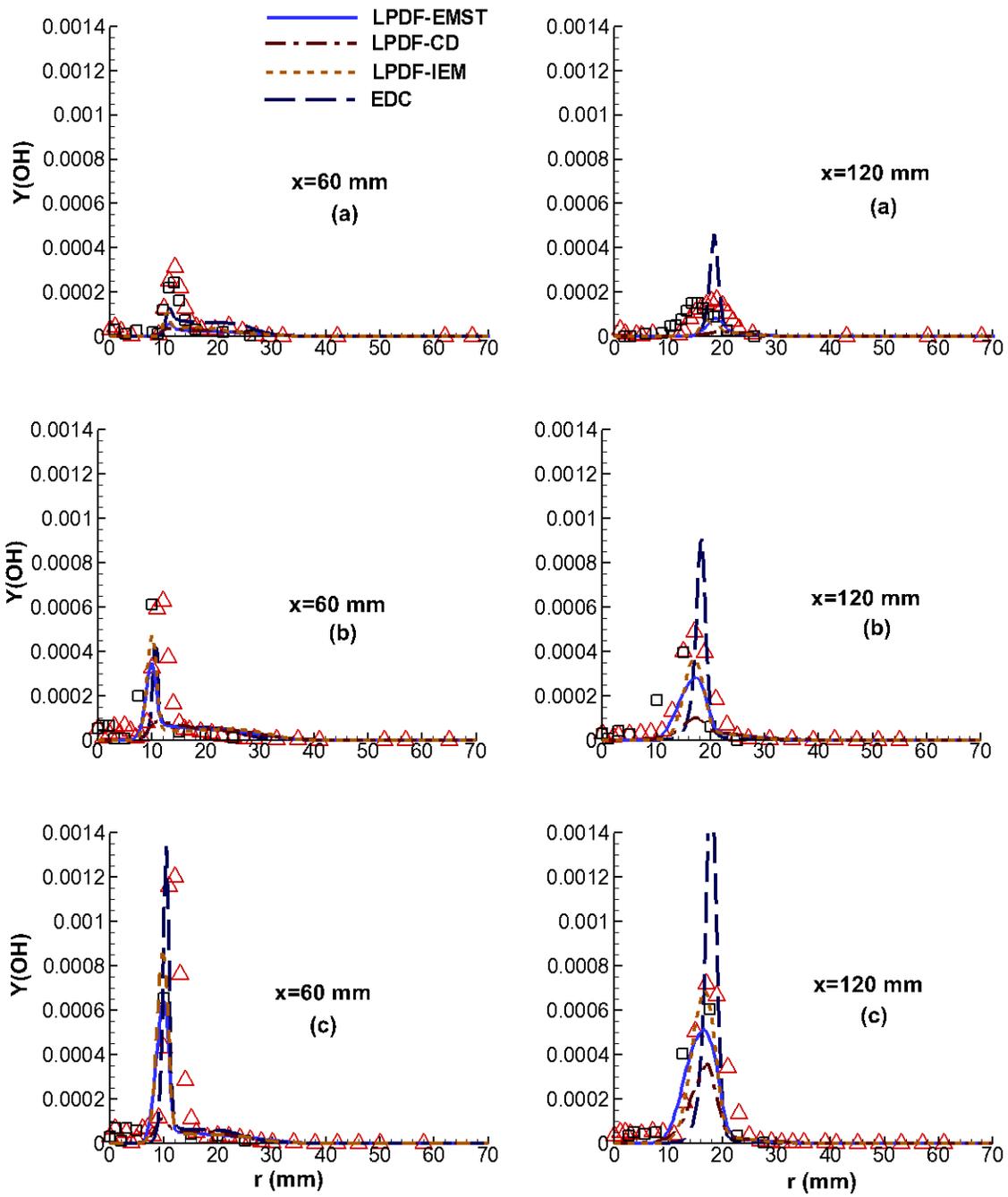

Figure 16

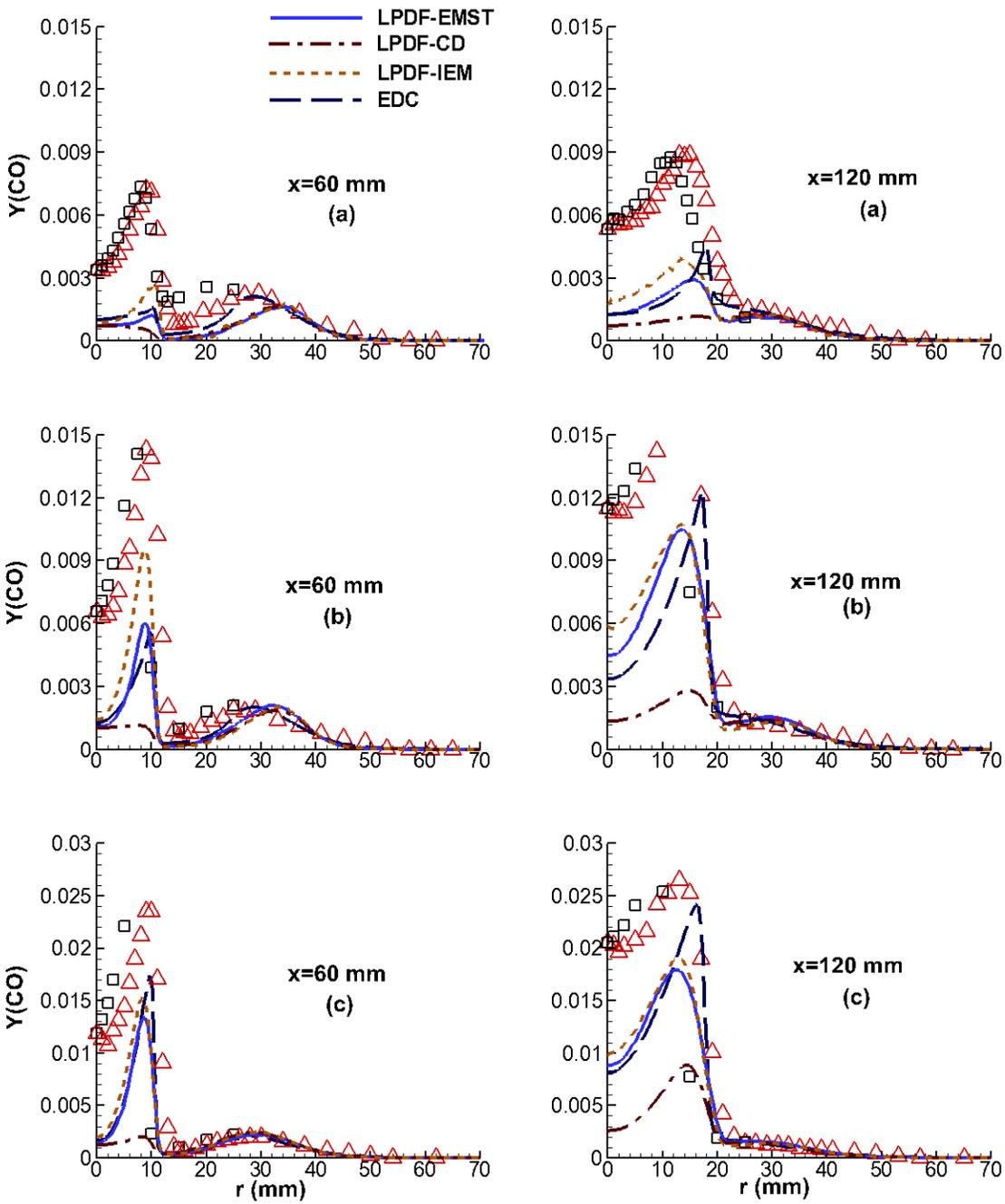

Figure 17

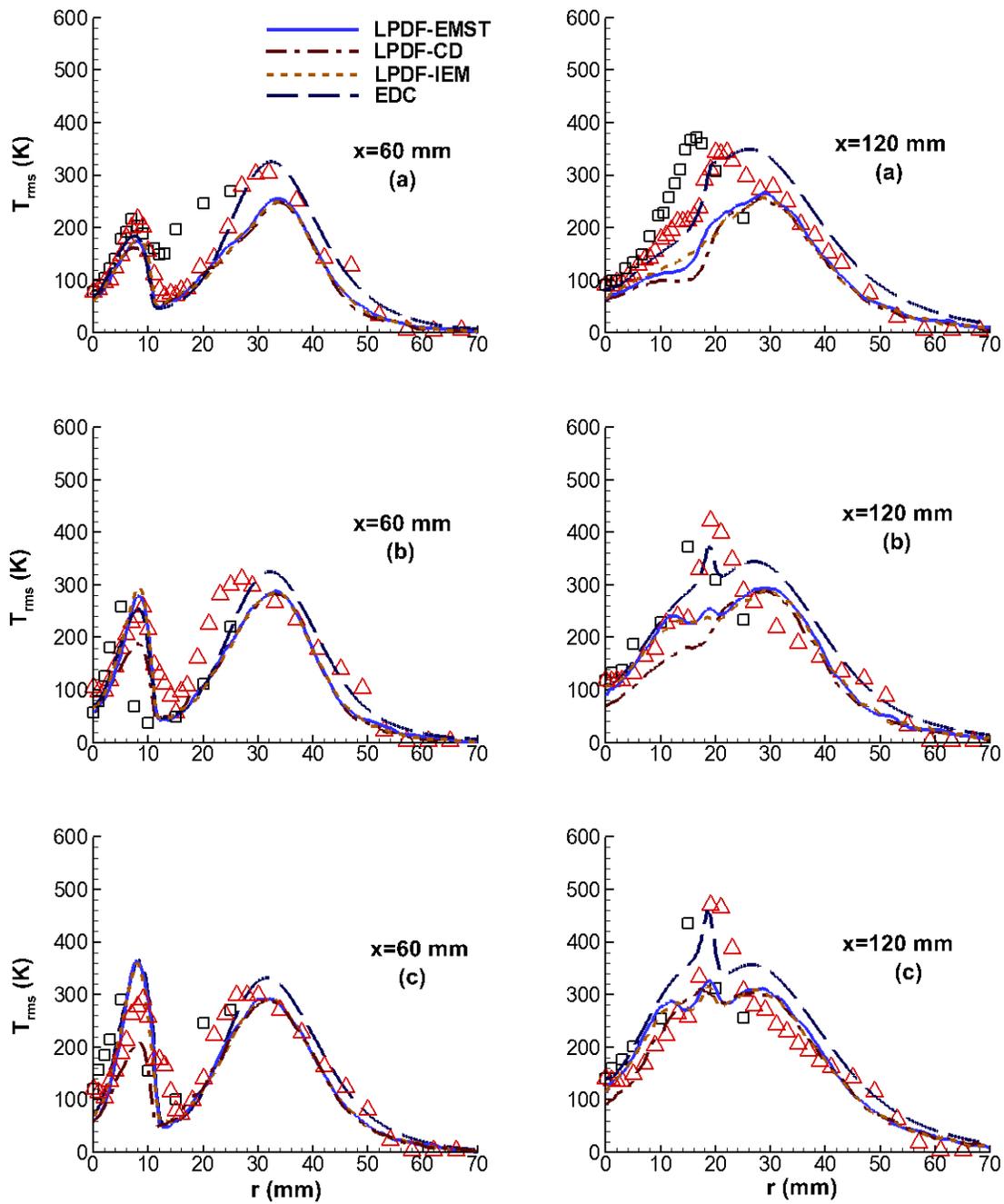

Figure 18

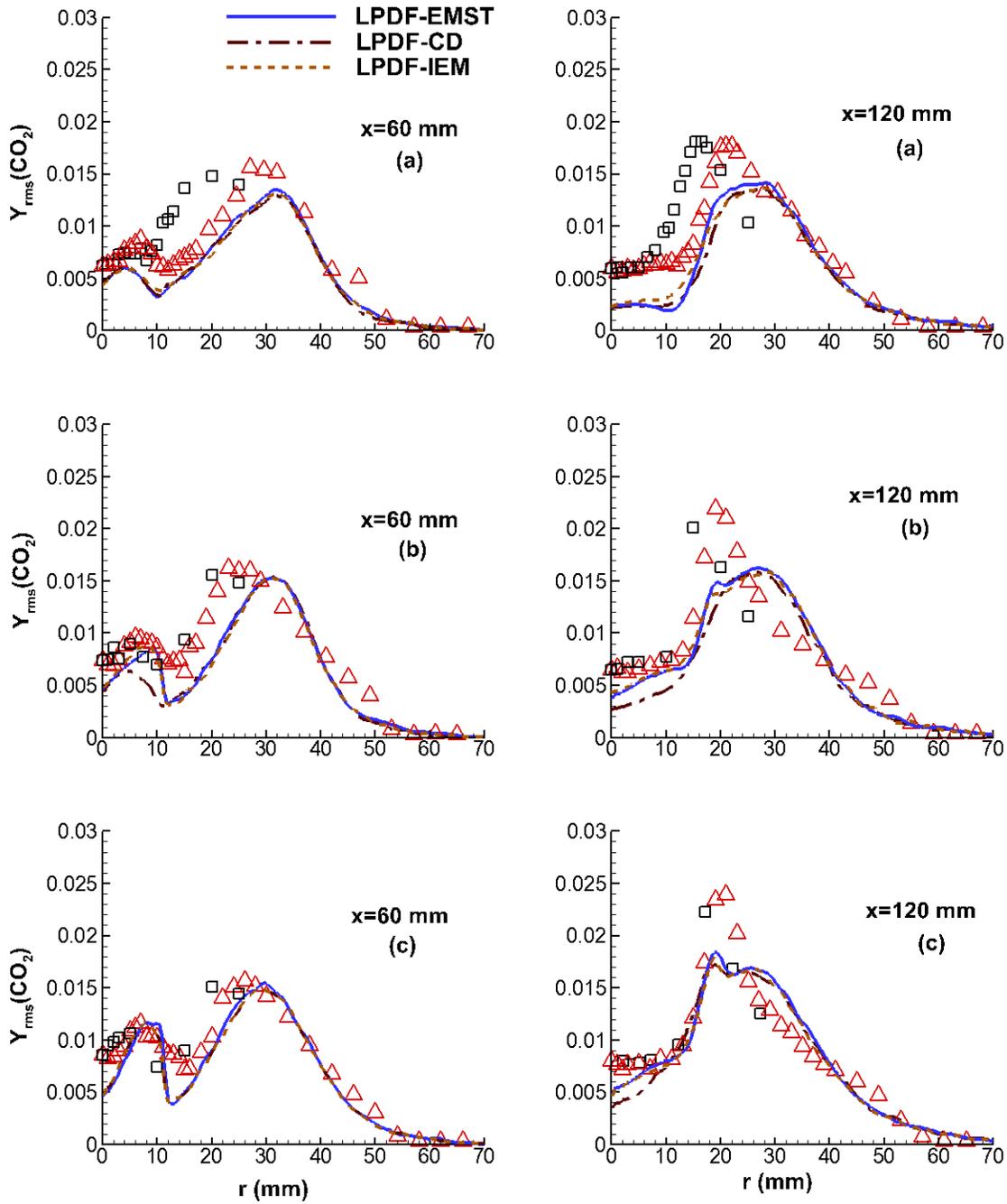

Figure 19

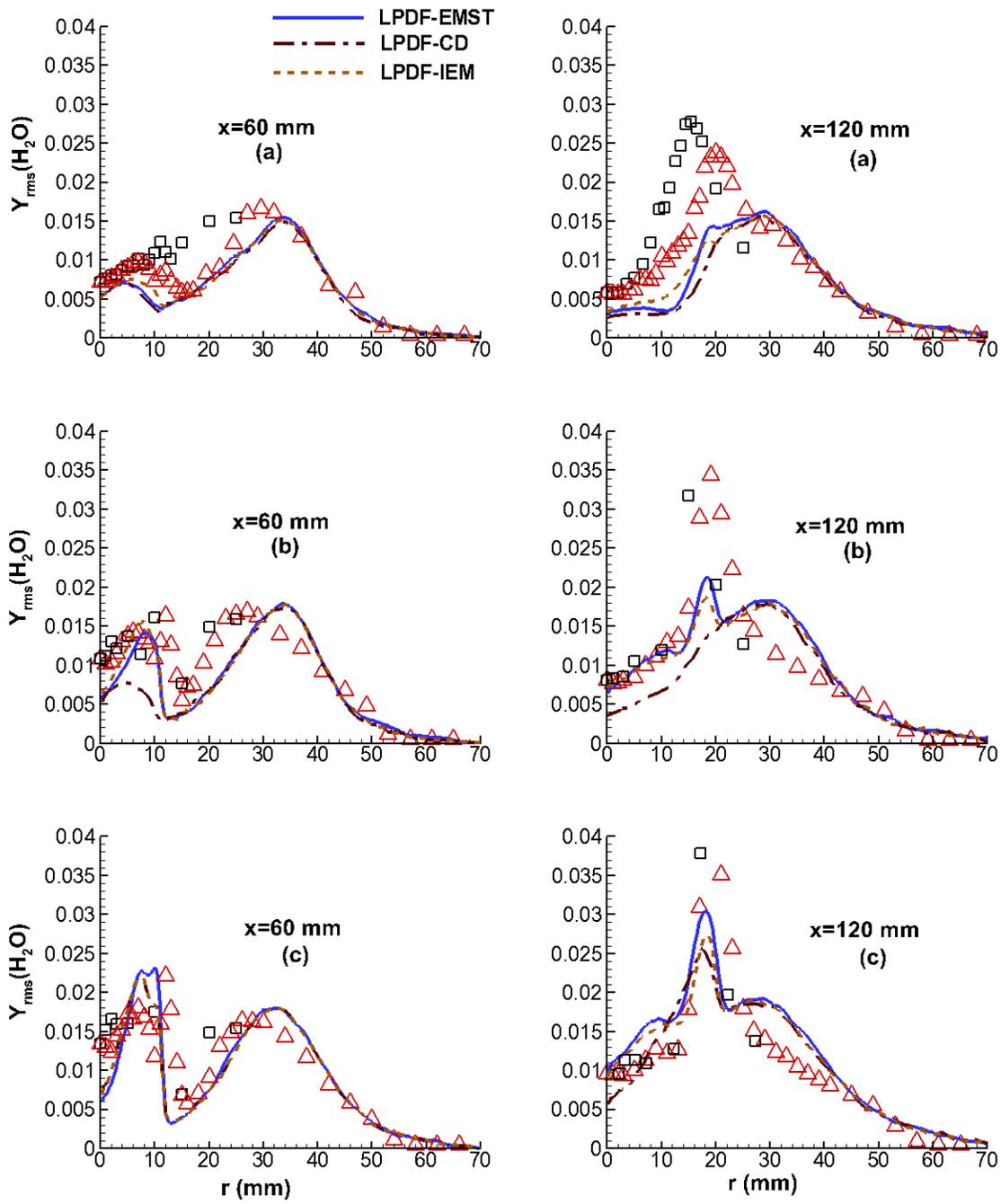

Figure 20

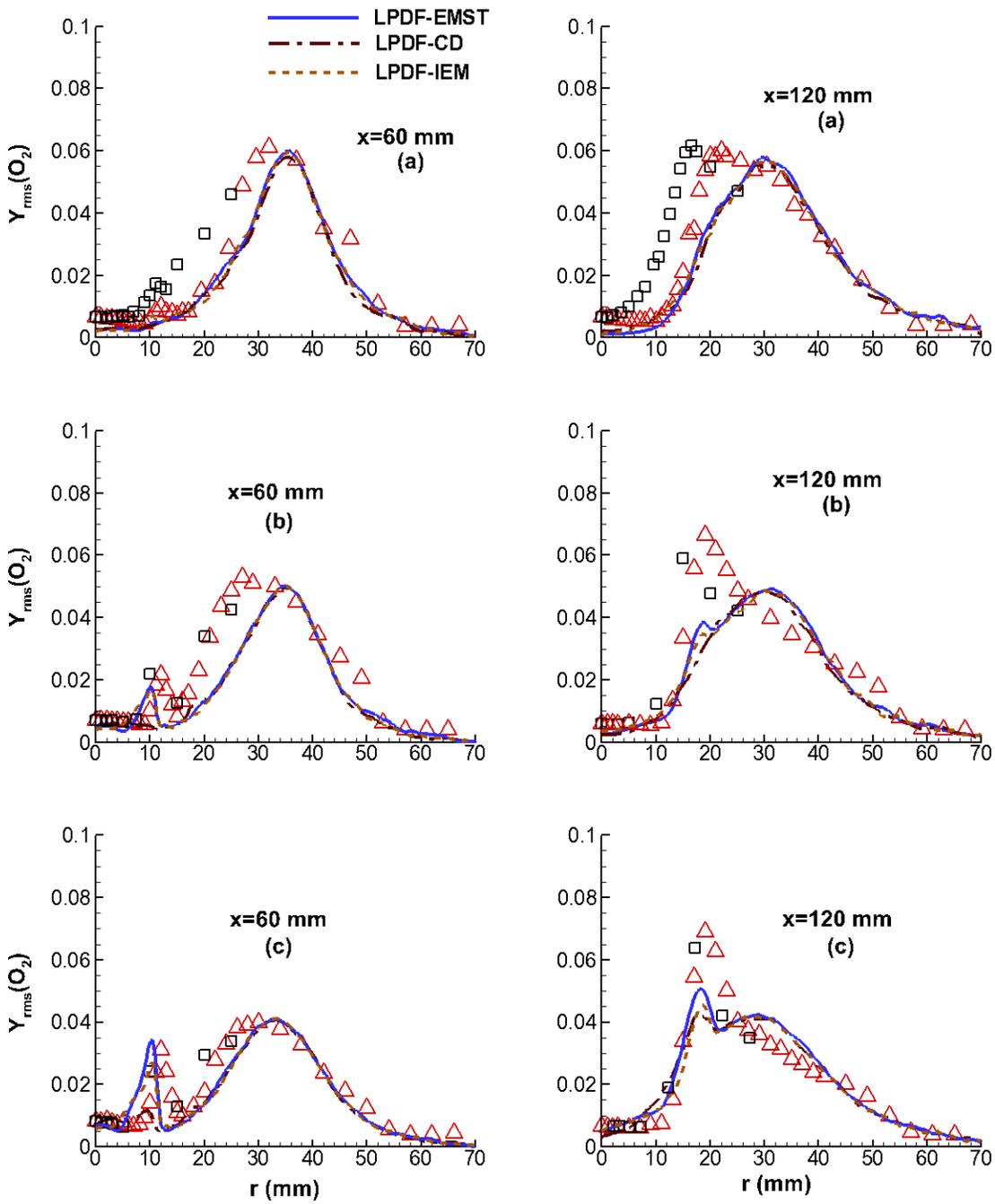

Figure 21

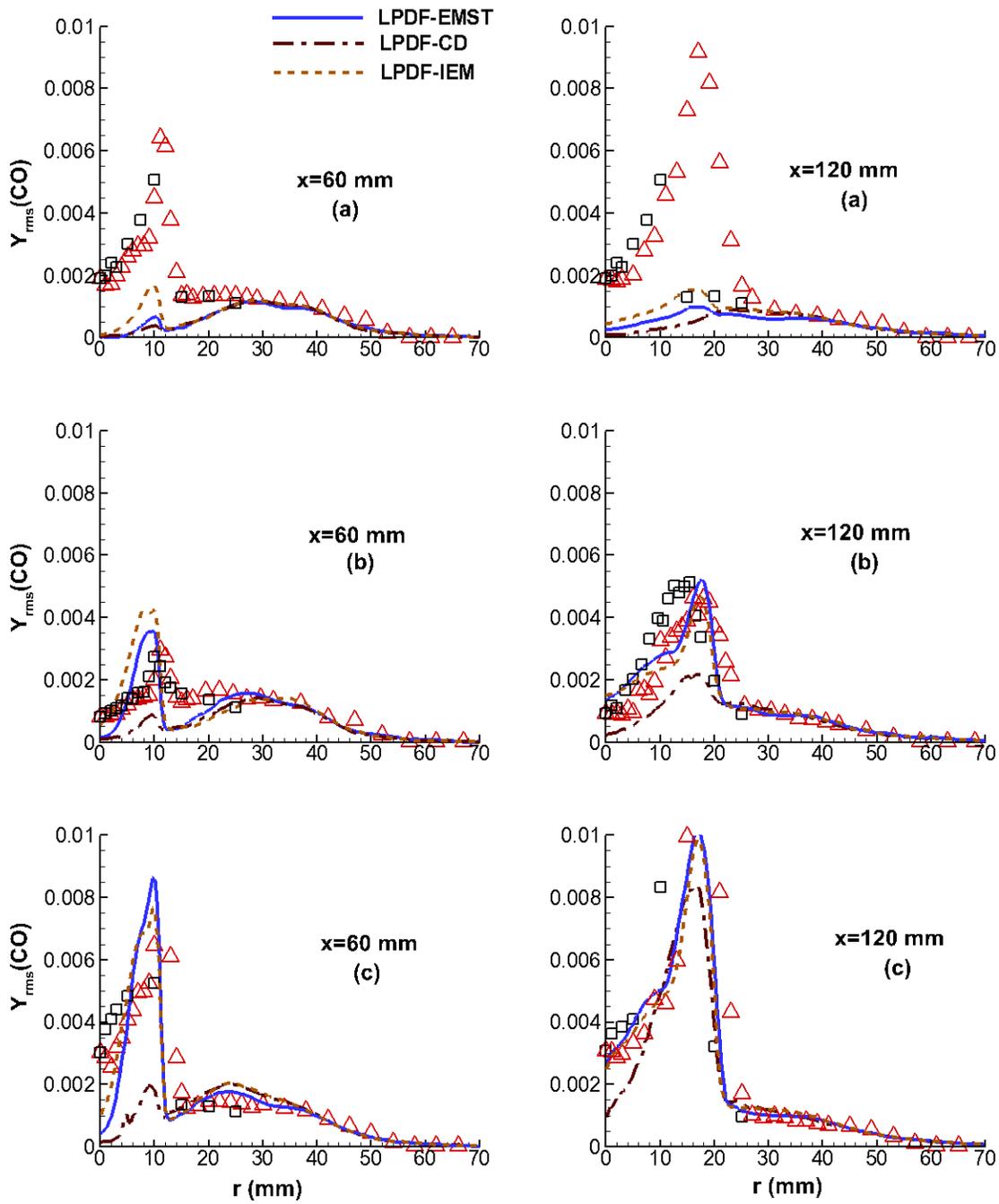

Figure 22

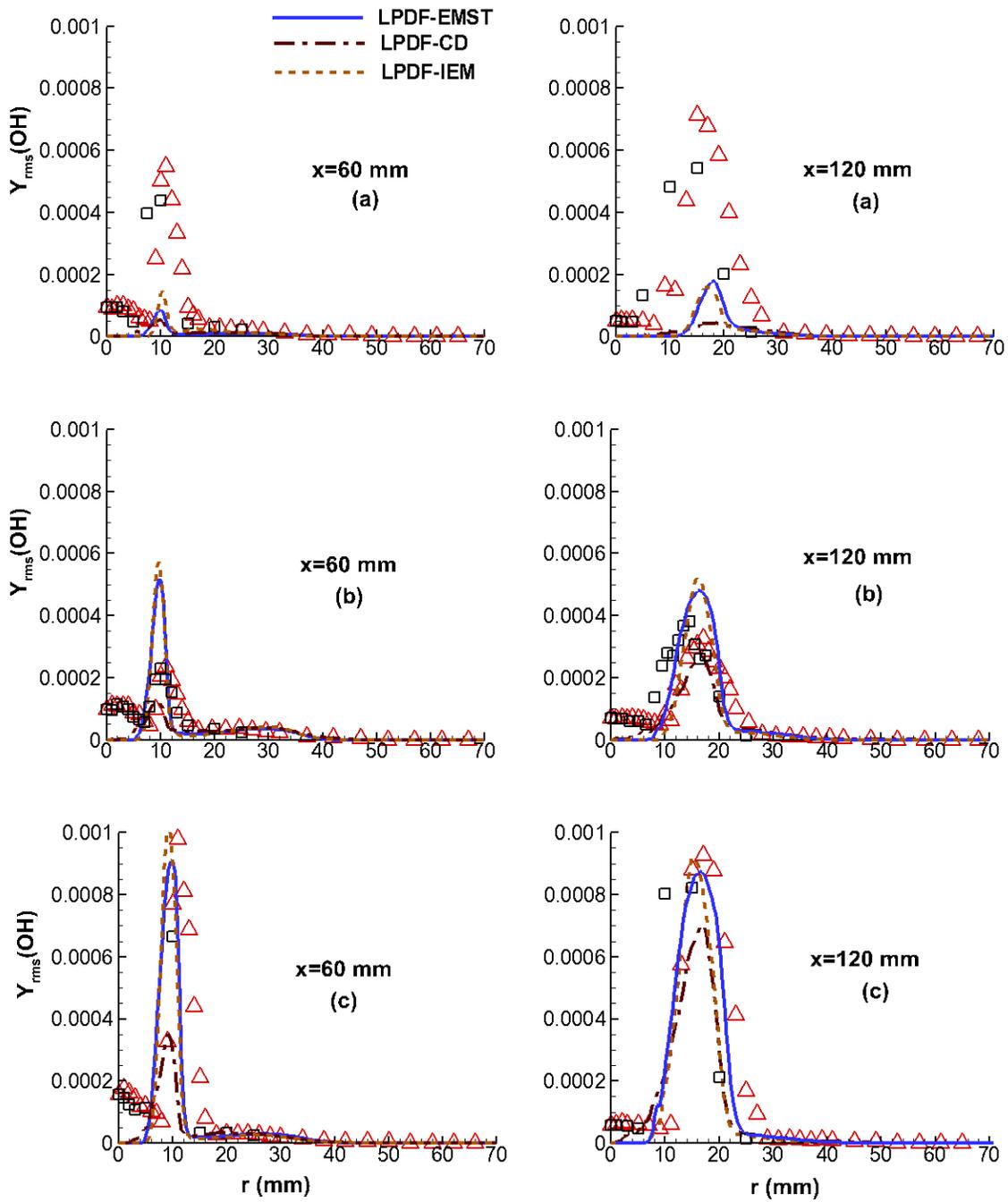

Figure 23

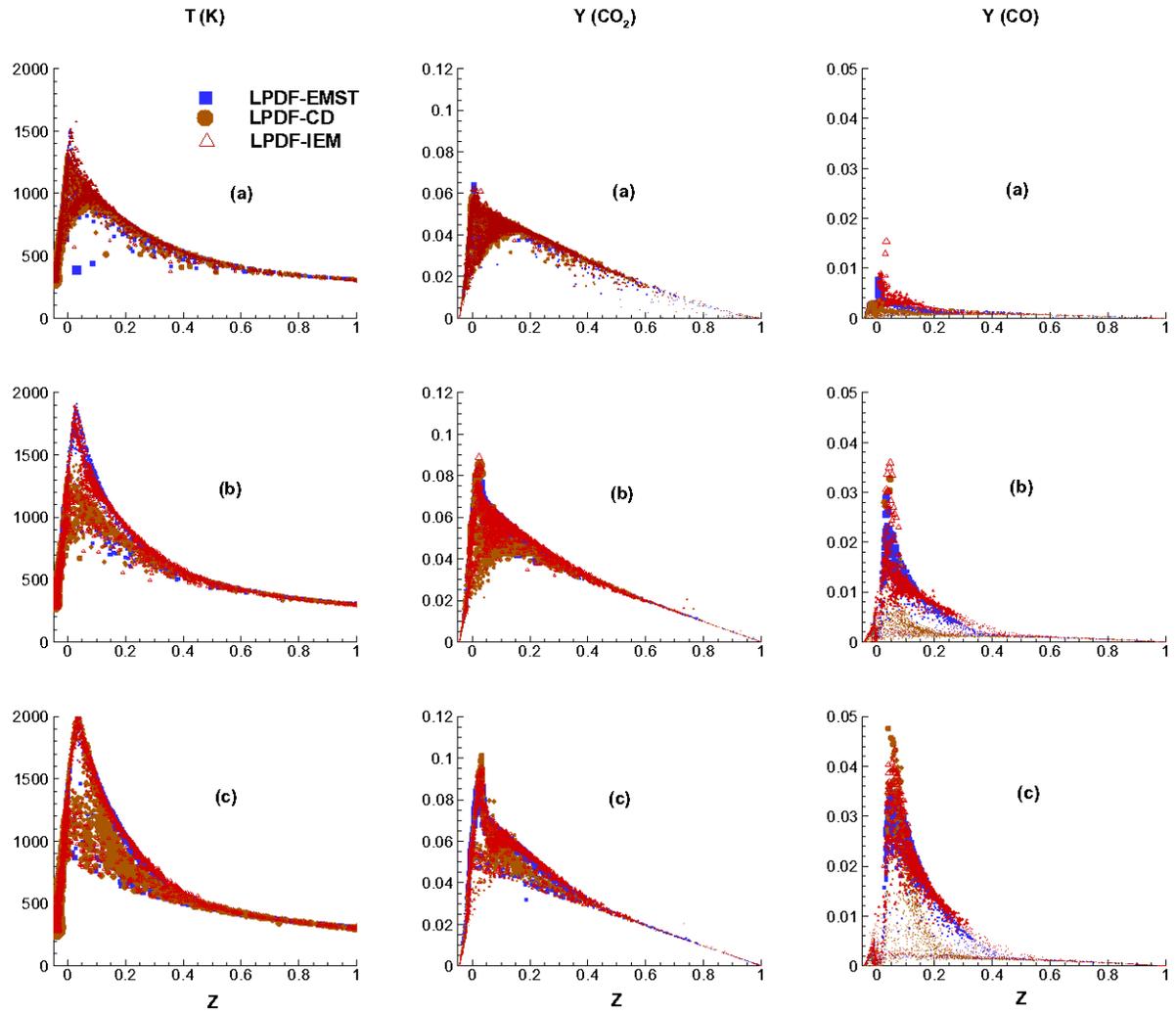

Figure 24

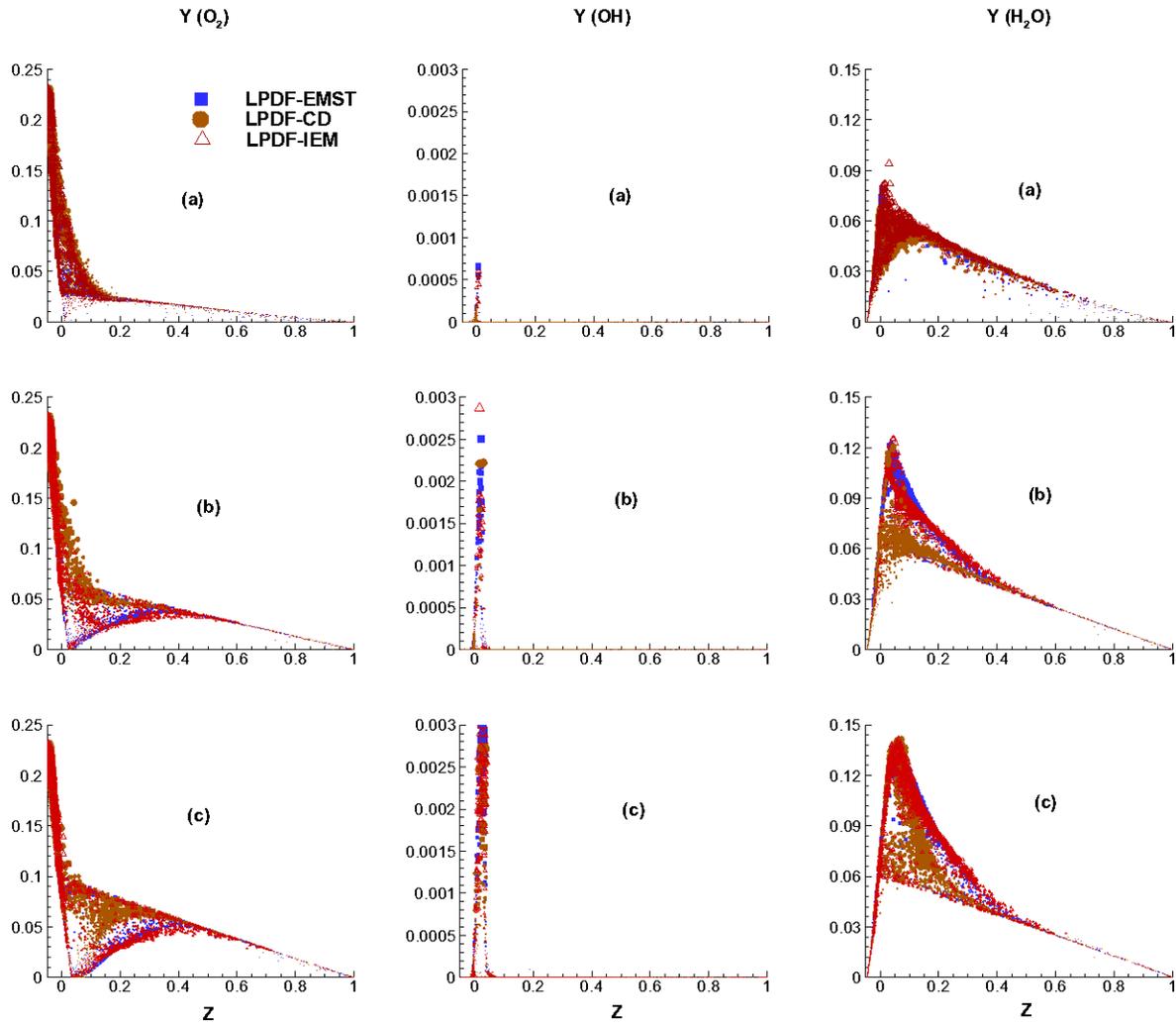

Figure 25